\documentclass[prd,aps,superscriptaddress,nofootinbib,amsmath,amssymb,showpacs,9pt]{revtex4}
\usepackage{graphicx}
\usepackage{braket}
\usepackage{nicefrac}
\usepackage{float}%
\usepackage{amsmath}
\usepackage{hyperref}
\usepackage{multirow}
\usepackage{slashed}
 \usepackage{bbold} 

\usepackage[usenames, dvipsnames]{color}

\def\slashchar#1{\setbox0=\hbox{$#1$}
   \dimen0=\wd0 \setbox1=\hbox{/} \dimen1=\wd1
   \ifdim\dimen0>\dimen1 \rlap{\hbox to \dimen0{\hfil/\hfil}} #1
   \else  \rlap{\hbox to \dimen1{\hfil$#1$\hfil}} / \fi}
\def\p{\slashchar{p}}
\def\q{\slashchar{q}}

\graphicspath{{./eps/}}

\begin{document}

\title{Charged current neutrino and antineutrino induced associated particle production from nucleons}

\author{A. \surname{Fatima}}
\affiliation{Department of Physics, Aligarh Muslim University, Aligarh-202 002, India}
\author{M. Sajjad \surname{Athar}}
\email{sajathar@gmail.com}
\affiliation{Department of Physics, Aligarh Muslim University, Aligarh-202 002, India}
\author{S. K. \surname{Singh}}
\affiliation{Department of Physics, Aligarh Muslim University, Aligarh-202 002, India}
\begin{abstract} 
In this work, we study the charged-current (anti)neutrino-induced associated particle~($K\Lambda$) production~($\Delta S=0$) from free nucleons in the energy region of a few GeV, relevant to the (anti)neutrino oscillation experiments with accelerator and atmospheric neutrinos. We employ a model based on effective Lagrangians to evaluate the contributions from the nonresonant and the resonant diagrams. The nonresonant background terms are calculated using a microscopic model derived from the SU(3) chiral Lagrangians. For the resonant contributions, we consider the low-lying spin-$\frac{1}{2}$ resonances, such as $S_{11}(1650)$, $P_{11}(1710)$, $P_{11}(1880)$, and $S_{11}(1895)$, and spin-$\frac{3}{2}$ resonances, such as $P_{13}(1720)$ and $P_{13}(1900)$, which have finite branching ratios to the $K\Lambda$ channel. These resonant contributions are modelled using an effective phenomenological Lagrangian approach, with strong couplings determined from the experimental branching ratios and the decay widths to the $K\Lambda$ channel. To fix the parameters of the vector current interaction, 
the model is first used to reproduce satisfactorily the MAMI$@$Mainz experimental data on the real photon induced scattering off the nucleon resulting an eta meson in the final state and with the CLAS$@$JLab data for the $K\Lambda$ production in the final state. 
The PCAC hypothesis and the generalized Goldberger-Treiman relation are used to fix the parameters of the axial vector interaction.
The model is then applied to study the weak production of $K\Lambda$ induced by the neutrinos and antineutrinos, and predicts the numerical values for the $Q^2$-distribution~$\left(\frac{d\sigma}{dQ^2}\right)$, the kinetic energy distribution for the outgoing kaon~$\left(\frac{d\sigma}{dp_K}\right)$, and the total scattering cross sections~($\sigma$) with and without a cut on the center of mass energy $W$. The results presented in this work are relevant for the present and future accelerator experiments like MicroBooNE, T2K, NOvA, MINERvA,  SBND, ICARUS, T2-HyperK and DUNE as well as for the atmospheric neutrino experiments. 
\end{abstract}
\maketitle

\section{Introduction}
\label{intro}
The discovery that the neutrinos have nonzero masses through the observation of the neutrino oscillation phenomenon is one of the rare clues pointing towards the new physics beyond the standard model~(BSM).
Consequently, the study of the phenomenology of the neutrino oscillation has led to the measurement of the various oscillation parameters like the three mixing angles including the measurement of very small but non-zero value of the angle $\theta_{13}$. 
However, no definite measurement of the CP violating phase angle $\delta$ has been reported, and the question of the neutrino mass hierarchy involving the measurements of the three mass square differences i.e., $\Delta m_{12}^{2}$, $\Delta m_{23}^{2}$, and $\Delta m_{13}^{2}$, is still not settled due to the lack of consensus on the values of these parameters obtained from the various experiments. 
In order to have more precise measurements of the mixing angles and the CP violating phase angle as well as to settle the issue of the neutrino mass hierarchy, various current and future neutrino experiments have been planned in laboratories around the world, which will also explore new physics beyond the standard model. 
Most of these experiments use neutrino beams in the energy region of a few GeV and the target material containing moderate to heavy nuclei. 
In this energy region, a reliable knowledge of the nuclear medium effects on the neutrino-nuclear cross sections as well as a precise understanding of the basic neutrino-nucleon cross sections from the free nucleon targets are essential inputs. 
 Whether the nuclear medium effects are experimentally constrained or theoretically derived from the first principles, the independent understanding of the elementary level neutrino-nucleon amplitudes for the various physical processes in the energy region of a few GeV is essential. 
 The study of the various neutrino-nucleon processes is not only vital for the analysis of the neutrino oscillation experiments but also for gaining deeper insights into the hadronic interactions in the weak  sector, where additional contribution from the axial vector current exists alongside the vector current contributions. Moreover, in the weak sector, the vector and axial vector form factors appearing in the matrix elements of the hadronic current are determined assuming  various symmetries like the time reversal invariance, SU(3) symmetry, conserved vector current~(CVC) hypothesis, partial conservation of the axial vector current~(PCAC) hypothesis, etc. Therefore, a better understanding of the neutrino-nucleon cross section is important in order to test these symmetry properties~\cite{SajjadAthar:2022pjt, Athar:2020kqn, SajjadAthar:2021prg, Ruso:2022qes}. For this, we need experimental measurements of neutrino-nucleon cross sections which has been recently highlighted in a white paper by the Snowmass collaboration~\cite{Alvarez-Ruso:2022ctb}. 
 
 In the few GeV energy region of (anti)neutrino,  the contribution to the cross section comes from various reaction channels such as quasielastic, inelastic, and deep inelastic scattering processes which have different energy dependence~\cite{SajjadAthar:2024etq}. 
 The inelastic scattering on the nucleons involves a wide range of processes triggered when (anti)neutrinos interact with the nucleon target. Among these, the single pion production is the most prominent inelastic reaction. However, nucleons illuminated by neutrinos can also radiate photons and, with increasing energy transfer, give rise to multiple pions, eta, kaons, and other mesons along with a baryon with or without strangeness. 
 In general, the strange mesons like kaons are produced either through the $\Delta S =0$ reaction along with a hyperon in the final state, or through the  $\Delta S = \pm 1$ reaction along with a nucleon in the final state.
The Cabibbo-suppressed (anti)neutrino–nucleon interactions that leads to the production of kaons through the $\Delta S = \pm 1$ processes requires lower threshold energy but has smaller cross sections and has been discussed earlier by us and others~\cite{RafiAlam:2010kf, Alam:2011vwg}.  In this work, we focus on the associated particle production of kaons induced by the $\Delta S=0$ reactions, which results in final states such as $Y K$, where $Y$ denotes hyperons like $\Lambda$, $\Sigma$, etc. and $K$ may be $K^+$ or $K^0$, and features larger cross sections but the threshold energy for their production is high. 

The history of studying weak interaction induced associated particle production processes is more than half century old, which started with the theoretical works of Shrock~\cite{Shrock:1975an} using Born approximation on the weak charged and neutral current induced $\Delta S = 0$ reactions off the nucleon targets and followed by many others like Mecklenburg~\cite{Mecklenburg:1976pk}, Amer~\cite{Amer:1977fy}, and Dewan~\cite{Dewan:1981ab}. In the last two-three decades, many theoretical efforts have been made to study the single pion production, the single kaon production, the single hyperon production, etc.~\cite{SajjadAthar:2022pjt} but only limited efforts have been put to study the associated particle production processes, like the works of Adera et al.~\cite{Adera:2010zz} using Born approximation and Nakamura et al.~\cite{Nakamura:2015rta} using the dynamical coupled channel~(DCC) model. 
{ Adera et al.~\cite{Adera:2010zz} have studied the differential cross sections~(at $d^4 \sigma$ level) in the threshold region using only the nucleon Born terms, where the hadronic current receives contribution from the s, t, and u channel diagrams and the contributions from the resonance excitations have not been considered. 
No results have been discussed at the level of the total scattering cross section $\sigma$ and $Q^2$ distribution. 
On the other hand, Nakamura et al.~\cite{Nakamura:2015rta} have used the DCC model to calculate the total scattering cross section for the neutrino induced production of pion, kaon, eta, and 2 pions in the final state. The hadronic current receives contribution from both the nonresonant Born terms as well as the resonance excitations. We have made a comparison of the total scattering cross section obtained in the present calculations with the available results of  Nakamura et al.~\cite{Nakamura:2015rta} for the kaon production through the associated particle production~($K\Lambda$) process~(Fig.~\ref{assoc-neut-compare}). We have also compared our present results with the earlier results obtained by Shrock~\cite{Shrock:1975an}.}
The process of associated particle production also plays an important role in deepening our understanding of the fundamental symmetries of the Standard Model (SM), the structure of weak hadronic form factors, the strange-quark content of the nucleon, etc. In addition to its theoretical significance, kaon production via the associated particle production processes also constitutes a critical background in proton decay searches, such as $p \rightarrow K \bar{\nu}$. Consequently, achieving a reliable and accurate estimation of the cross sections for these background-contributing processes to the proton decay channels is of paramount importance and has been strongly emphasized in the literature~\cite{Solomey:2005rs, Mann:1986ht}. 

Experimentally, the observations of neutrino-induced associated particle production were carried out in earlier decades at ANL~\cite{Barish:1974ye, Barish:1978pj}, CERN~\cite{Deden:1975pa, Erriquez:1978pg}, BNL~\cite{Baker:1981tx, Baker:1986xx}, and Fermilab~\cite{Son:1983xh}. However, these measurements were limited by very low statistics and significant systematic uncertainties.   The theoretical estimates significantly underestimate the experimental data, in some cases by almost a factor of four~\cite{Datchev:2002}. For example, the theoretical results presented by Shrock~\cite{Shrock:1975an} and Mecklenburg~\cite{Mecklenburg:1976pk}, averaged over the ANL flux, show strong disagreement with the experimentally measured value, and there is also a significant difference between the predictions from these two theoretical calculations.  Modern neutrino experiments, equipped with high-intensity (anti)neutrino beams, are now in a better position to revisit and refine these studies.  
 Nonetheless, the cross sections used in the commonly adopted neutrino event generators, such as GENIE~\cite{GENIE:2021npt}, NEUT~\cite{Hayato:2021heg}, etc., account only for the resonant associated kaon production based on the Rein–Sehgal model, which was originally developed for pion production~\cite{Rein:1980wg}.
 This stark discrepancy highlights the urgent need for improved theoretical models and more accurate cross section predictions for the weak interaction induced associated particle production. Such advancements are essential not only for the precision neutrino physics but also for enhancing the sensitivity of rare event searches, including proton decay.

 In the current generation of accelerator experiments, there are no experimental data on the specific reaction channels involving strangeness changing~($\Delta S =1$) and strangeness conserving~($\Delta S=0$) processes.
 The recent experimental measurements on the charged current induced kaon production by the MINERvA~\cite{MINERvA:2016zyp} and MicroBooNE~\cite{MicroBooNE:2025kqo} collaborations are focussed on the inclusive kaon production, 
 where a $K^+$ meson along with a muon are observed in the final state, while the other hadrons remain undetected. 
 The MINERvA collaboration~\cite{MINERvA:2016ymg} has also reported the neutral current induced inclusive kaon production. The MINERvA collaboration~\cite{MINERvA:2016ymg, MINERvA:2016zyp} has reported the results for the kinetic energy distribution of kaons from the plastic scintillator target, which are consistent with the GENIE simulation results especially in the case of neutral current induced $K^+$ production.  
 The MicroBooNE collaboration~\cite{MicroBooNE:2025kqo} has  reported the results for the flux integrated cross section of the charged current neutrino induced inclusive $K^{+}$ production from the argon target, and compared the results with the different neutrino event generators like GENIE, NEUT, and NuWro. 
 The results are consistent with the predictions from the different neutrino generators but large uncertainties in the experimental results require more precise data so that the interaction mechanism may be theoretically understood. 
 However, exclusive production of $K^+$ with or without a hyperon in the final state has not been reported in any of the recent measurements.

 In this work, we present a theoretical calculation to obtain the differential and total scattering cross sections for $\Lambda K$ production for the $\nu_\mu(\bar\nu_\mu)$ induced scattering off the nucleon target. The weak hadronic current receives contribution from the nonresonant Born terms as well as from the resonance excitations. 
The Born terms are obtained using an SU(3) symmetric chiral model. 
The contribution from the resonance terms is considered from the excitation of both, the spin $\frac{1}{2}$ resonances like $S_{11} (1650)$, $P_{11} (1710)$, $P_{11}(1880)$, $S_{11}(1895)$ and the spin $\frac{3}{2}$ resonances like 
$P_{13} (1720)$ and $P_{13} (1900)$.  
In the case of spin $\frac{1}{2}$ resonances, the $s$-channel hadronic currents for the positive and 
negative parity resonances with the explicit form of 
the vector and axial-vector form factors for the isospin $\frac{1}{2}$ resonances have been used.  
The vector form factors of the nucleon-resonance~($N-R$) transition are expressed in terms of the electromagnetic $N-R$ form factors using the isospin symmetry, which in turn are determined from the helicity amplitudes of the resonance excitations. 

Recently the treatment of spin $\frac{3}{2}$ resonances has been discussed in literature using Rarita-Schwinger formalism~\cite{Rarita:1941mf} as well as Pascalutsa formalism~\cite{Pascalutsa:1999zz}. It is well known that the Rarita–Schwinger formalism is not unique for
describing the spin $\frac{3}{2}$ field (as well as for the higher spin fields) and has a problem
associated with the lower spin degrees of freedom. This leads to some ambiguities in
describing the propagation of the off-shell spin $\frac{3}{2}$ field using a propagator specially
in the presence of external interactions like the electromagnetic and/or the strong interactions. The
problem has been discussed extensively in literature for many years ever since the
field theory of higher spins was developed using either the vector-spinor formalism~\cite{Rarita:1941mf}
or the multi-spinor formalism~\cite{Fierz}. Consequently, there are various prescriptions
for treating the propagator and the effective Lagrangians for the interacting
fields of higher spin in a consistent way for describing the interaction of spin $\frac{3}{2}$
fields~\cite{Pascalutsa:1994tp}. One of the most popular prescriptions given Pascalutsa and Timmermans~\cite{Pascalutsa:1999zz}
has been investigated further in the latest works of Mart~\cite{Mart:2019jtb} and Vrancx et al.~\cite{Vrancx:2011qv}
and many other references cited there. We find~\cite{Fatima:2025} that the experimental results for the photon induced associated particle production off the nucleon better explain the experimental data when the numerical calculations are done by including the spin $\frac{3}{2}$ resonance contribution using the formalism developed by Pascalutsa and Timmermans~\cite{Pascalutsa:1999zz}. Therefore, all the (anti)neutrino results presented in this work are obtained using the Pascalutsa and Timmermans~\cite{Pascalutsa:1999zz} formalism  when spin-$\frac{3}{2}$-resonant states are included.

In view of the above discussion, the parameters of the vector current interaction in the case of (anti)neutrino-induced production of $K\Lambda$ are determined by fitting the experimental data on the electromagnetic production of $K\Lambda$ and $\eta$ meson.
In the case of photoproduction of $\eta$, our model explains the MAMI experimental data~\cite{CrystalBallatMAMI:2010slt, A2:2014pie}, where spin-$\frac{1}{2}$ resonances and nonresonant background terms are considered. Then, the experimental data from the CLAS collaboration~\cite{CLAS05} for the photon-induced associated particle production ($K\Lambda$) are explained, where additional resonances, including
spin-$\frac{3}{2}$ resonances, are also taken into account. The $Q^2$ dependence of the vector form factors are fixed by explaining the CLAS electroproduction data~\cite{Denizli:2007tq} for the virtual photon-induced eta production. The PCAC
hypothesis and the generalized Goldberger-Treiman~(GT) relation are used to fix the parameters of the axial-vector
current interaction.

Accordingly, in Secs.~\ref{photo} and \ref{electro}, we present the formalism for the photon and electron induced associated particle production, respectively, and compared them with the experimental results. The formalism for the charged current (anti)neutrino induced associated particle production is
presented in Sec.~\ref{neutrino}. The results and discussions are presented in
Sec.~\ref{results}. In Sec.~\ref{summary}, we conclude our findings.

\section{Photo- and Electro- production of $K\Lambda$ and $N\eta$}
\subsection{Photoproduction}\label{photo}
The differential cross section for the photoproduction of associated kaons and $\eta$ off the proton target, {  i.e.},
\begin{eqnarray}
 \label{eq:KLambda}
 \gamma(q) + p(p ) &\longrightarrow& \Lambda (p^{\prime}) + K^+(p_{K}), \\
 \label{eq:eta}
 \gamma(q) + p(p ) &\longrightarrow& p (p^{\prime}) + \eta(p_{\eta}), 
\end{eqnarray}
is written as~\cite{Fatima:2022nfn}:
\begin{equation}\label{dsig}
\left. \frac{d \sigma}{d \Omega}\right|_{CM} = \frac{1}{64 \pi^{2} s} \frac{|\vec{p}\;^{\prime}|}{|\vec{p}\;|} 
\overline{\sum_{r}} \sum_{spin} |\mathcal{M}^{r}|^2,
\end{equation}
where the quantities in the parentheses of Eqs.~(\ref{eq:KLambda}) and (\ref{eq:eta}) represent the four momenta of the 
corresponding particles. The CM energy $\sqrt{s}$ is expressed as $
 s = W^2 = (q + p)^2 = M^{2} + 2M E_{\gamma} ,$
with $E_{\gamma}$ being the energy of the incoming photon in the laboratory frame. 
$ \overline{\sum} \sum | \mathcal M^{r} |^2$ is the square of the transition matrix element $\mathcal{M}^{r}$, for the photon 
polarization state $r$, averaged and summed over the initial and final spin states, where $\mathcal{M}^{r}$ is written in terms of 
the real photon polarization vector $\epsilon_{\mu}^{r}$, as
\begin{equation}
\mathcal{M}^{r} = e \epsilon_{\mu}^{r} (q) J^{\mu} ,
\end{equation}
with $e$ being the electromagnetic coupling constant and $J^{\mu} $ being the matrix element
of the electromagnetic current taken between the hadronic initial and final states. The hadronic matrix element  receives contribution from the nonresonant Born terms, kaon resonances being exchanged in the $t$-channel diagrams, and the terms corresponding to 
the resonance excitations and their subsequent decay to $K\Lambda$ or $N\eta$ mode. 
The hadronic currents for the nonresonant Born terms are 
 obtained using the nonlinear sigma model and the total hadronic current $J^{\mu}$ is  obtained by adding the currents corresponding to the nonresonant and resonance terms coherently. For the detailed description of the formalism, readers are referred to Ref.~\cite{SajjadAthar:2022pjt}. The detailed formalism for the photon and electron induced $\eta$ production, along with the relevant Feynman diagrams, hadronic currents, etc., are discussed in Refs.~\cite{Fatima:2022nfn, Fatima:2023fez}. 
 
    \begin{figure}
 \begin{center}
    \includegraphics[height=3cm,width=3.9cm]{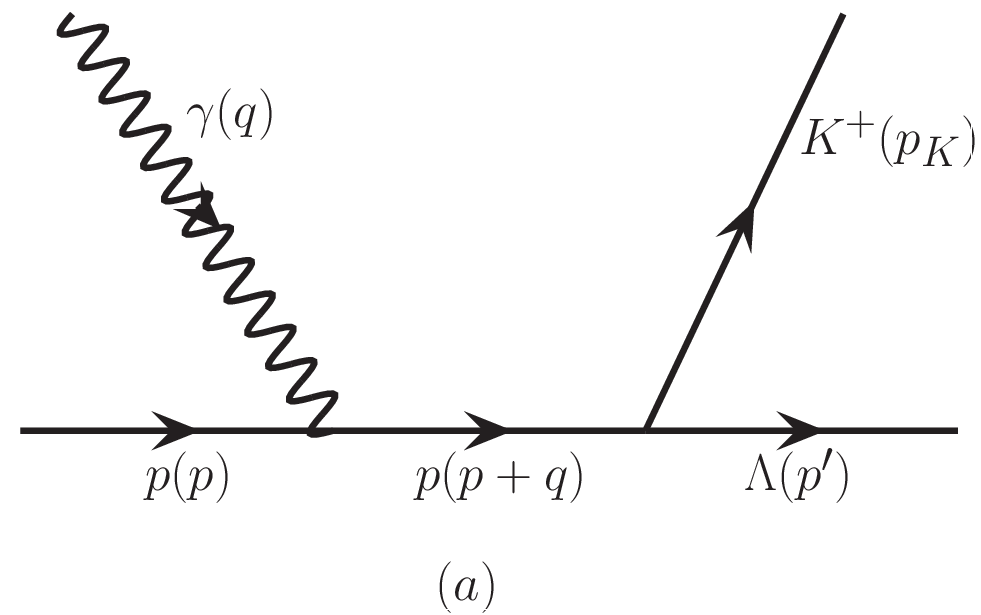}
    \hspace{5mm}
    \includegraphics[height=3.5cm,width=3.3cm]{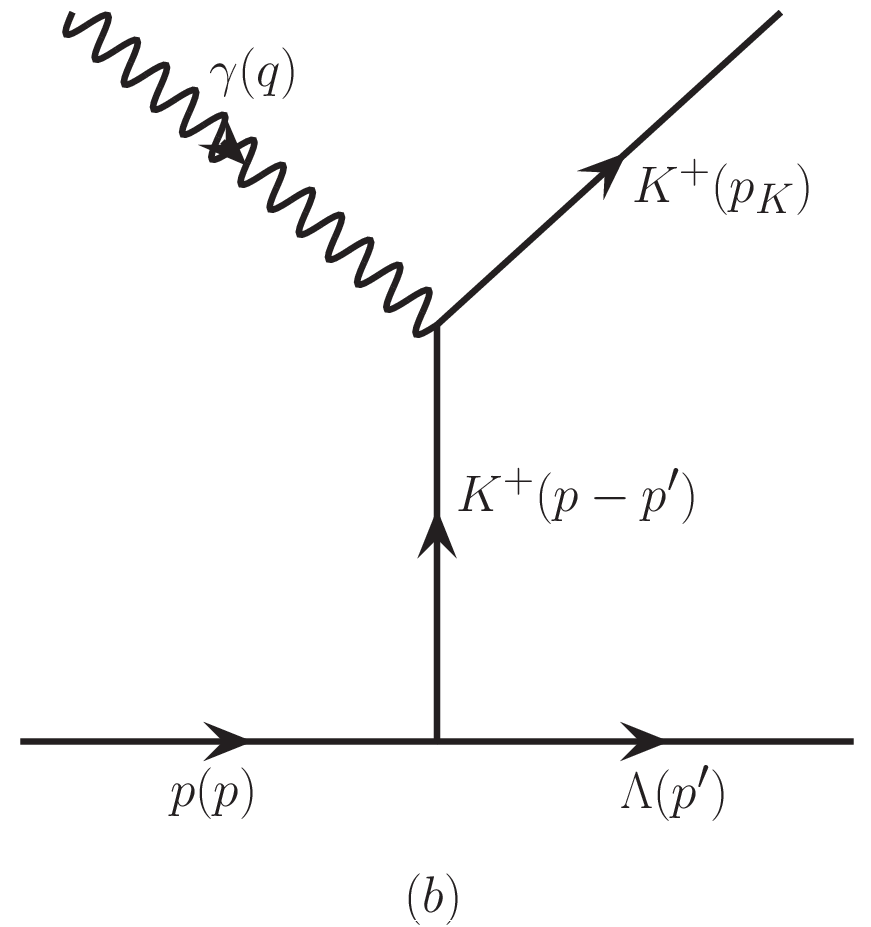}
    \hspace{5mm}
    \includegraphics[height=3cm,width=3.9cm]{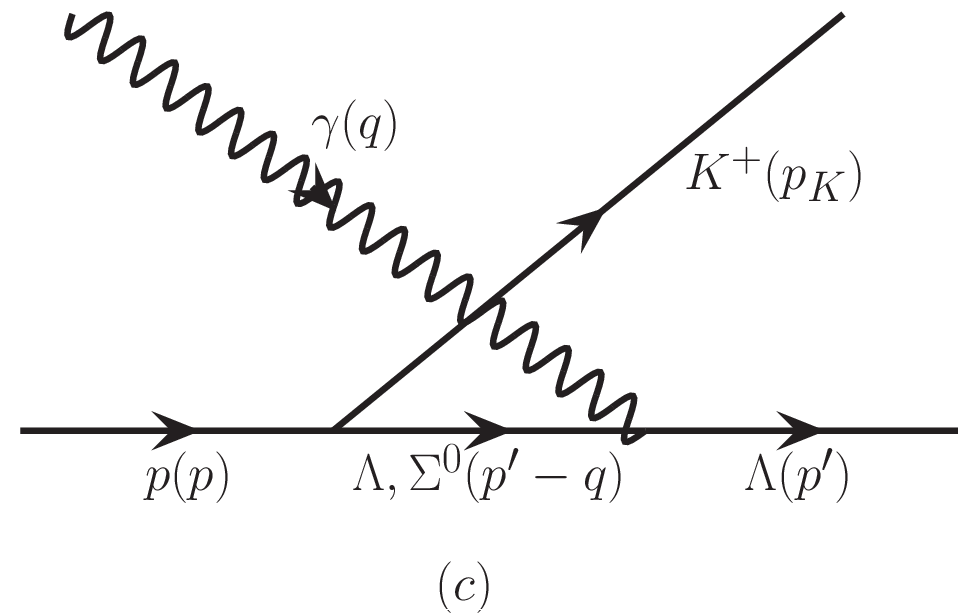}
    \hspace{5mm}    
    \includegraphics[height=3cm,width=4cm]{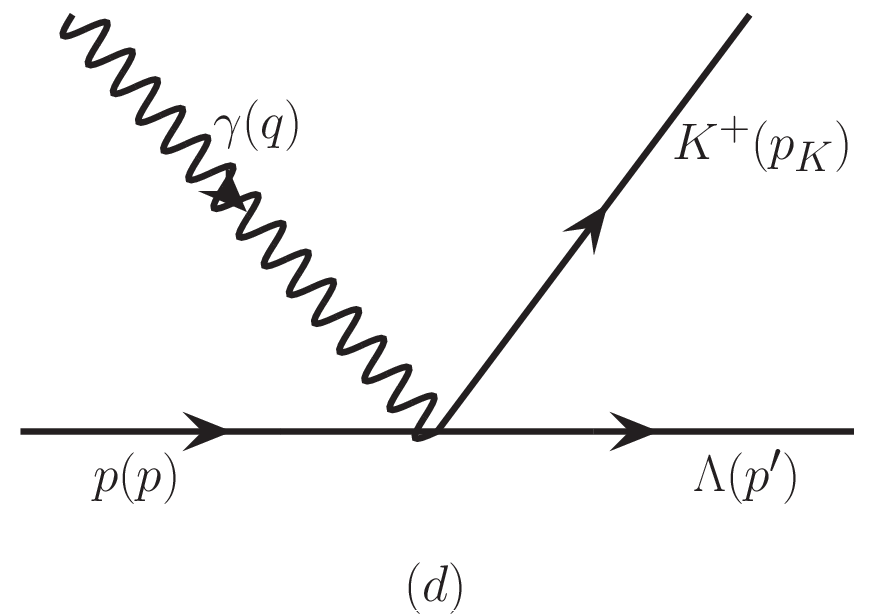}
    \vspace{5mm}
    
    \includegraphics[height=3.5cm,width=3.9cm]{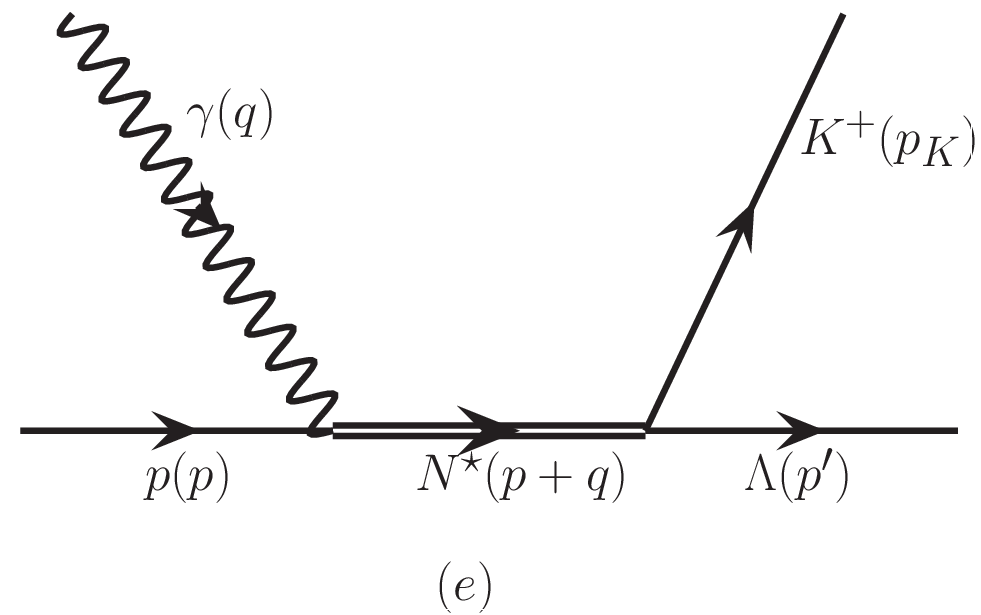}
    \hspace{5mm}
    \includegraphics[height=3cm,width=3.3cm]{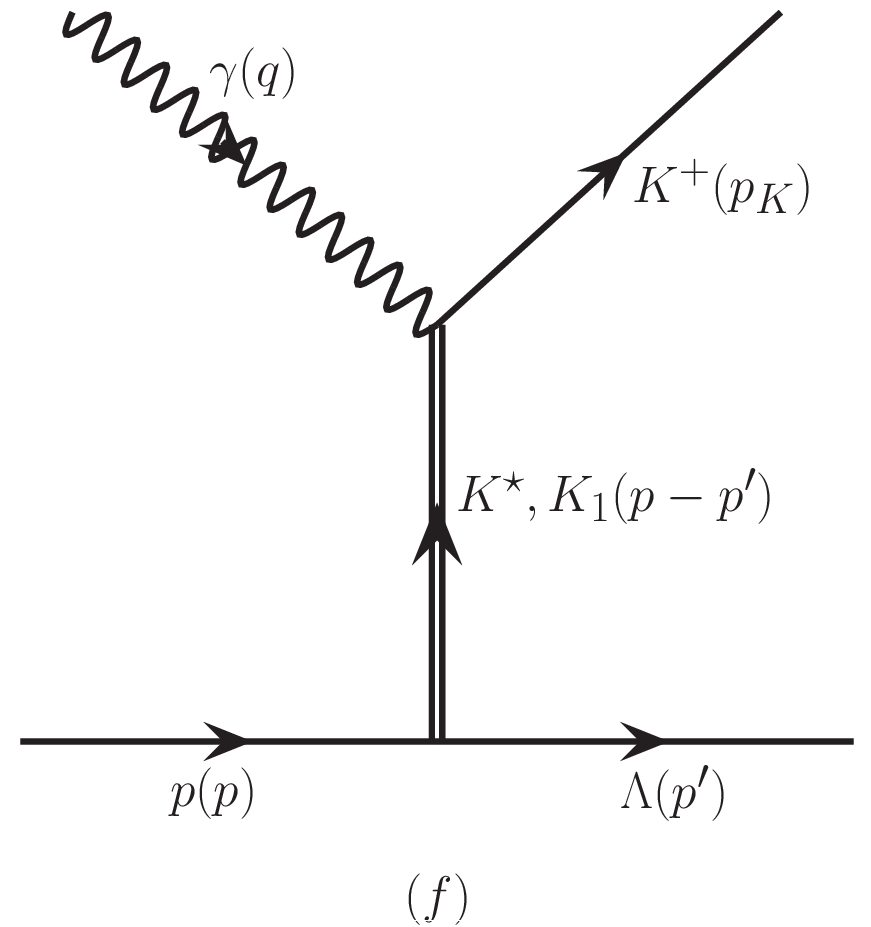}
  \caption{Feynman diagram for the various channels possible for the process $ \gamma (q) + p(p) \rightarrow K^{+}(p_{k}) + 
  \Lambda(p^{\prime})$. (a)~$s$-channel, (b)~$t$-channel, (c)~$u$-channel and (d)~contact term constitute the non-resonant 
  terms. (e)~nucleon resonances in the $s$-channel, and (f)~kaon resonances in the $t$-channel. The quantities in the bracket represent four momenta of the corresponding 
  particles.}\label{fyn_dia}
   \end{center}
 \end{figure}
 
The hadronic currents for the various non-resonant terms contributing to the $K\Lambda$ production, which are shown diagrammatically in Fig.~\ref{fyn_dia}, are expressed as~\cite{Fatima:2020tyh}:
\begin{eqnarray}\label{j:s}
J^\mu \arrowvert_{s} &=&ie A_{s}~F_{s}(s) \bar u(p^\prime) \slashed{p}_k \gamma_5 \frac{ \slashed{p} + \slashed{q} + M}
  {s -M^2} \left(\gamma^\mu e_{p} +i \frac{\kappa_{p}}{2 M} \sigma^{\mu \nu} q_\nu \right) u(p), \\
  \label{j:t}
  J^\mu \arrowvert_{t}&=& ie A_{t}~ F_{t}(t) \bar u(p^\prime)\left[(\slashed{p} - \slashed{p}^{\prime}) \cdot \gamma_{5} \right] 
  u(p) \frac{(2 p_{k}^{\mu} - q^{\mu})}{t - M_{k}^{2}} , \\
  \label{j:ulam}
J^\mu \arrowvert_{u \Lambda} &=&ie A_{u}^{\Lambda} ~F_{u}^{\Lambda} (u) \bar u(p^\prime) \left(\gamma^\mu e_{\Lambda} + i 
\frac{\kappa_{\Lambda}}{2 M_{\Lambda}} \sigma^{\mu \nu} q_\nu \right) \frac{ \slashed{p}^{\prime} -\slashed{q} + M_{\Lambda}}
{u - M_{\Lambda}^2} \slashed{p}_k \gamma_5 u(p), \\
\label{j:usig}
J^\mu \arrowvert_{u \Sigma^{0}} &=&ie A_{u}^{\Sigma^{0}} ~F_{u}^{\Sigma^{0}} (u) \bar u(p^\prime) \left(\gamma^\mu 
e_{\Sigma^{0}} + i \frac{\kappa_{\Sigma^{0}}}{2 M_{\Sigma^{0}}} \sigma^{\mu \nu} q_\nu \right) 
\frac{\slashed{p}^{\prime} -\slashed{q} + M_{\Sigma^{0}}} {u -M_{\Sigma^{0}}^2} \slashed{p}_k \gamma_5 u(p), \\
\label{j:CT}
J^\mu \arrowvert_{CT} &=&-i e A_{CT} ~F_{CT} \bar u(p^\prime) \; \gamma^\mu \gamma_5 \; u(p),
\end{eqnarray}
where $CT$ stands for the contact term and $s,~t,~u$ are the Mandelstam variables defined as
\begin{eqnarray}
 s = (p+q)^2, \qquad \qquad \qquad t= (p - p^{\prime})^{2}, \qquad \qquad \qquad u = (p^{\prime} - q)^{2}.
\end{eqnarray}
 $A_{i}$'s; $i=s,t,u,CT$ are the coupling strengths of $s$-, $t$-, $u$- channels and the 
contact term, respectively, and are obtained as
\begin{eqnarray}\label{eq:coupling}
 A_{s} = A_{t} = A_{u}^{\Lambda} = A_{CT} &=& - \left(\frac{D + 3F}{2 \sqrt{3} f_{K}}\right)  ; \qquad \quad
 A_{u}^{\Sigma^{0}} = 
 \left(\frac{D-F}{2 f_{K}} \right).
\end{eqnarray}
All these couplings of the non-resonant terms are generated by the chiral symmetry and are fixed by the low energy electroweak 
phenomenology consistent with the experimental data.
$D$ and $F$ are the axial-vector couplings of the baryon octet and $f_{K}=105$~MeV~\cite{Faessler:2008ix} is the kaon decay constant. 

In order to take into account the hadronic structure of the nucleons, the form factors $F_s(s)$, $F_t(t)$, $F_u(u)$ and $F_{CT}(s,t,u)$, are 
introduced at the strong vertices. { We use a dipole form for the hadronic form factor}~\cite{Fatima:2020tyh}:
\begin{equation}\label{FF_Born}
F_{x} (x) = \frac{\Lambda_{B}^{4}}{\Lambda_{B}^{4} + (x - M^{2})^{2}}, \qquad \qquad \quad x=s,t,u
\end{equation}
where $\Lambda_{B}$ is the cut-off parameter taken to be the same for the $s$-, $t$- and $u$- channel nonresonant Born terms, and $x$ represents the Mandelstam variables $s,~t,~u$. The value of $\Lambda_{B}$ 
is fitted to the experimental data and the best fitted value is 
$\Lambda_{B}=0.54$~GeV.  For the strong form factor of the contact term $F_{CT}(s,t,u)$, we have used the prescription of Davidson and Workman~\cite{DW}, who parameterize $F_{CT}$ in terms of $F_{s}$ and $F_{t}$ as:
\begin{equation}
 F_{CT}(s,t,u) = F_{s}(s) + F_{t}(t) - F_{s}(s) \times F_{t}(t).
\end{equation}
One of the most important property of the electromagnetic current is the
gauge invariance which corresponds to the current conservation. 
In Ref.~\cite{Fatima:2020tyh}, we have discussed in detail the implementation of the gauge invariance in the case of the photoproduction of $K\Lambda$ for the Born terms while for the spin $\frac{1}{2}$ and $\frac{3}{2}$ nucleon resonances, the hadronic current is gauge invariant by construction. 

\begin{table*}
\centering
\begin{tabular*}{180mm}{@{\extracolsep{\fill}}ccccccccc}\hline \hline
\multicolumn{2}{c}{Resonance $\rightarrow$} & \multirow{2}{*}{$S_{11}(1535)$} & \multirow{2}{*}{$S_{11}(1650)$} & \multirow{2}{*}{$P_{11} (1710)$} &  \multirow{2}{*}{$P_{13} (1720)$} & \multirow{2}{*}{$P_{11} (1880)$} &\multirow{2}{*}{$S_{11} (1895)$} &  \multirow{2}{*}{$P_{13} (1900)$}\\

\multicolumn{2}{c}{Parameters $\downarrow$} &&& &&&&\\ \hline

\multicolumn{2}{c}{$M_{R}$ (GeV)} & $1.510 \pm 0.01$ & $1.655 \pm 0.015$ & $1.700 \pm 0.02$ & $1.680 \pm 0.02$ & $1.860 \pm 0.04$ & $1.910\pm 0.02$ & $1.920\pm 0.02$ \\ \hline

\multicolumn{2}{c}{$\Gamma_{R}$ (GeV)} & $0.130 \pm 0.02$ & $0.135 \pm 0.035$ & $0.120 \pm 0.04$ & $0.150 \pm 0.05$ &  $0.230 \pm 0.05$ & $0.110 \pm 0.03$ & $0.130 \pm 0.03$ \\ \hline

\multicolumn{2}{c}{$I(J^P)$} &$\frac{1}{2}(\frac{1}{2}^{-})$& $\frac{1}{2}(\frac{1}{2}^{-})$ & $\frac{1}{2}(\frac{1}{2}^{+})$ &$\frac{1}{2}(\frac{3}{2}^{+})$ & $\frac{1}{2}(\frac{1}{2}^{+})$ & $\frac{1}{2}(\frac{1}{2}^{-})$ &$\frac{1}{2}(\frac{3}{2}^{+})$ \\ \hline 

\multirow{4}{*}{BR (in \%)} & $N\pi$ & $32-52$~(43)& $50-70$~(60) &
$5-20$~(16) & $8-14$~(11)& $3-31$~(34) & $2-18$~(23) &  $1-20$~(12) \\ 

& $N\eta$ & $30-55$~(40) & $15-35$~(25) & $10-50$~(20) & $1-5$~(3)& $1-55$~(20) & $15-45$~(30) &$2-14$~(8) \\ 

 &$K\Lambda$ &$-$& $5-15$~(10) & $5-25$~(15) & $4-19$~(2)& $1-3$~(2) & $3-23$~(13)& $2-20$~(4) \\ 

& $N\pi\pi$ &$4-31$~(17)& $20-58$~(5)& $14-48$~(49)&$>50$~(84) & $>32$~(44) & $17-74$~(34)& $>56$~(76) \\ \hline

\multicolumn{2}{c}{$|g_{RN\pi}|$} & 0.1019 & 0.0915 & 0.0418 & 0.1105 & 0.0466 & 0.0229& 0.0644 \\ \hline

\multicolumn{2}{c}{$|g_{RK\Lambda}|$} & -- & 0.1362 & 0.1973 & 1.29 & 0.0604& 0.08556& 0.5684\\ \hline \hline
\end{tabular*}
\caption{Properties of the spin $\frac{1}{2}$ resonances available in the PDG~\cite{ParticleDataGroup:2020ssz}, with Breit-Wigner mass $M_{R}$, the total decay width $\Gamma_{R}$,
isospin $I$, spin $J$, parity $P$, the branching ratio full range available from PDG~(used in the present calculations) into different meson-baryon channels like $N\pi$, $N\eta$, $K\Lambda$, and $N\pi\pi$, and the strong coupling constant $g_{RN\pi}$ and $g_{RK\Lambda}$.}
\label{tab:param-p2}
\end{table*}

\begin{table*}
\centering
\begin{tabular*}{160mm}{@{\extracolsep{\fill}}ccc c c c c  c c c}\hline\hline
&Resonance & Helicity amplitude & \multicolumn{3}{c}{ Proton target } & \multicolumn{3}{c}{ Neutron target } &\\ \hline
&&& ${\cal A}_{\alpha} (0)$ & $a_1$ & $b_{1}$ & ${\cal A}_{\alpha} (0)$ & $a_1$ & $b_{1}$ &\\ \hline
&\multirow{2}{*}{$S_{11}(1535)$} & $A_{\frac{1}{2}}$ & 95.0 & 0.85 & 0.85 & $-78.0$ & 1.75 & 1.75& \\ 
&&$ S_{\frac{1}{2}}$ & $-2.0$ & 1.9 & 0.81 & $32.5$ & 0.4 & 1.0& \\ \hline
&\multirow{2}{*}{$S_{11}(1650)$} & $A_{\frac{1}{2}}$ & 33.3 & 0.45 & 0.72 & $26.0$ & 0.1 & 2.5& \\ 
&&$ S_{\frac{1}{2}}$ & $2.5$ & 1.88 & 0.96 & $3.8$ & 0.4 & 0.71& \\ \hline
&\multirow{2}{*}{$P_{11}(1710)$} & $A_{\frac{1}{2}}$ & 55.0 & 1.0 & 1.05 & $-45.0$ & $-0.02$ & 0.95& \\ 
&&$ S_{\frac{1}{2}}$ & $4.4$ & 2.18 & 0.88 & $-31.5$ & 0.35 & 0.85& \\ \hline

&\multirow{3}{*}{$P_{13}(1720)$} & $A_{\frac{1}{2}}$ & 100.0 & 1.89 & 1.55 & $-2.9$ & $1.7$ & 1.55& \\ 
&&$ A_{\frac{3}{2}}$ & $-11.0$ & 10.0 & 1.55 & $-31.0$ & 3.0 & 1.55& \\ 
&&$ S_{\frac{1}{2}}$ & $-53.0$ & 2.46 & 1.55 & $0$ & 0 & 0& \\ \hline

&\multirow{2}{*}{$P_{11}(1880)$} & $A_{\frac{1}{2}}$ & $-60.0$ & 0.4 & 1.0 & $-45.0$ & $-0.02$ & 0.95& \\ 
&&$ S_{\frac{1}{2}}$ & $0.4$ & 0.75 & 0.5 & $-31.5$ & 0.35 & 0.85& \\ \hline
&\multirow{2}{*}{$S_{11}(1895)$} & $A_{\frac{1}{2}}$ & $-15.0$ & 1.45 & 0.6 & $26.0$ & $0.1$ & 2.5& \\ 
&&$ S_{\frac{1}{2}}$ & $-3.5$ & 0.88 & 0.6 & $3.8$ & 0.4 & 0.71& \\
\hline 
&\multirow{3}{*}{$P_{13}(1900)$} & $A_{\frac{1}{2}}$ & 8.0 & 1.89 & 1.55 & $-2.9$ & $12.7$ & 1.55& \\ 
&&$ A_{\frac{3}{2}}$ & $-98.0$ & 1.0 & 1.55 & $-31.0$ & 3.0 & 1.55& \\ 
&&$ S_{\frac{1}{2}}$ & $-10.0$ & 0.46 & 1.0 & $0$ & 0 & 0& \\ \hline
\hline
\end{tabular*}
\caption{Parameterization of the helicity amplitude for $S_{11} (1535)$, $S_{11}(1650)$, $P_{11}(1710)$, $P_{13}(1720)$, $P_{11}(1880)$, $S_{11}(1895)$, and $P_{13}(1900)$ resonances on the proton and neutron targets.  ${\cal A}_{\alpha} (0)$ is given in units of $10^{-3}$ GeV$^{-2}$ and the coefficients $a_1$  and $b_1$ in units of GeV$^{-2}$.}
\label{tab:resonance}
\end{table*}

We have taken into account those resonances, which have mass $M_{R}<2$~GeV and a significant branching ratio to the $K\Lambda$ decay mode reported in PDG~\cite{ParticleDataGroup:2020ssz}. Specifically, we have considered four spin $\frac{1}{2}$ resonances {  viz.} $S_{11} (1650)$, $P_{11} (1710)$, $P_{11}(1880)$, and $S_{11}(1895)$; and two spin $\frac{3}{2}$ resonances {  viz.} $P_{13}(1720)$ and $P_{13}(1900)$. The general properties of these resonances like mass, decay width, spin, etc. are given in Table~\ref{tab:param-p2}.

The most general form of the hadronic current for the $s$-channel processes where a resonance state $R$ 
is produced and decays to a $\Lambda$ baryon and a $K$ meson in the final state, is written as~\cite{Athar:2020kqn}:
\begin{eqnarray}
j^\mu\big|_{s}&=& F_{s}^{*} (s) ~\frac{g_{RK\Lambda}}{f_{K}} \bar u({p}\,') 
 \p_{K} \gamma_5 \Gamma_{s} \left( \frac{\p+\q+M_{R}}{s-M_{R}^2+ iM_{R} \Gamma_{R}}\right) 
 \Gamma^\mu_{\frac12 
 \pm} u({p}\,), 
\end{eqnarray}
where $\Gamma_{R}$ and $M_{R}$, respectively, are the decay width and mass of the resonance. $\Gamma_{s} = 1(\gamma_{5})$ stands for the positive~(negative) 
parity resonances, and $g_{RK\Lambda}$ is the strong coupling strength of the $ R K\Lambda$ vertex, which has been determined using the partial decay width of the resonance to $K\Lambda$ mode where the central values of the full width tabulated in Table~\ref{tab:param-p2} are used in the numerical calculations. The values of the strong coupling constant of the different resonances are also tabulated in Table~\ref{tab:param-p2}. The vertex functions $\Gamma_{\frac{1}{2}^{+}}^\mu$ and $\Gamma_{\frac{1}{2}^{-}}^\mu$ for the 
positive and negative parity resonances are defined as
\begin{align}\label{eq:vec_half_pos_EM}
  \Gamma^{\mu}_{\frac{1}{2}^\pm} &= {V}^{\mu}_\frac{1}{2}\Gamma_{s},
  \end{align}
where $V^{\mu}_{\frac{1}{2}}$ represents the vector current parameterized in terms of $F_{2}^{R^{+},R^{0}} $, as
 \begin{align}\label{eq:vectorspinhalf1}
  V^{\mu}_{\frac{1}{2}} & =\left[\frac{F_2^{R^{+},R^{0}}}{2 M} 
  i \sigma^{\mu\alpha} q_\alpha \right].
\end{align}
The coupling $F^{R^{+},R^0}_{2}$ is derived from the helicity amplitudes extracted from the real photon scattering 
experiments. The explicit relation between the coupling $F_2^{R^{+},R^0}$ and the helicity amplitude $A_{\frac{1}{2}}^{p,n}$ is given 
by~\cite{Fatima:2022nfn}:
\begin{eqnarray}\label{eq:hel_spin_12}
A_\frac{1}{2}^{p,n}&=& \sqrt{\frac{2 \pi \alpha}{M} \frac{(M_R \mp M)^2}{M_R^2 - M^2}} \left[ \frac{M_R \pm M}{2 M} F_2^{R^{+},R^0} 
\right] ,
\end{eqnarray}
where the upper~(lower) sign stands for the positive~(negative) parity resonance. $R^{+}$ and $R^{0}$ correspond, respectively, to the charged and neutral states of the isospin $\frac{1}{2}$ resonances. 
The value of the helicity amplitude $A_{\frac{1}{2}}^{p,n}$ for the different resonances are  quoted in 
Table~\ref{tab:resonance}.

In analogy with the nonresonant terms,  we have considered the following form factor at the strong 
vertex for the various resonances, in order to take into account their hadronic structure:
\begin{equation}\label{strong_FF_res}
F^{*}_{x} (x) = \frac{\Lambda_{R}^{4}}{\Lambda_{R}^{4} + (x - M_{R}^{2})^{2}}, \qquad \qquad x = s,t
\end{equation}
where $\Lambda_{R}$ is the cut-off parameter whose value is fitted to the experimental data and the best fit for $\Lambda_{R}$ for spin $\frac{1}{2}$ resonances is obtained as $\Lambda_{R} = 1$~GeV. 

To determine the value of the strong $RK\Lambda$ coupling, we start by writing the most general form of 
$RK\Lambda$ interaction Lagrangian~\cite{Fatima:2022nfn}:
\begin{align}\label{eq:spin12_lag}
 \mathcal{L}_{R K\Lambda}(x) &= \frac{g_{ R K\Lambda} }{f_{K}}\bar{\Psi}_{R}(x) \; 
 \Gamma^{\mu}_{s} \;
  \partial_\mu K(x) \,\Psi(x) ,
\end{align}
where $\Psi(x)$ is the nucleon field, ${\Psi}_{R}(x)$ 
is the resonance field, and $K(x)$ is the kaon field. The interaction vertex $\Gamma^{\mu}_{s} = \gamma^\mu \gamma^5$~($\gamma^\mu$) stands for the positive~(negative) parity 
resonance states. 

Using the above Lagrangian, one obtains the expression for the decay width in the resonance rest frame as~\cite{SajjadAthar:2022pjt}:
\begin{align}\label{eq:12_width}
 \Gamma_{R \rightarrow K\Lambda} &= \frac{\mathcal{C}}{4\pi} \left(\frac{g_{RK\Lambda }}{f_{K}}
 \right)^2 \left(M_R \pm M\right)^2 \frac{E_{N} \mp M}{M_R} |\vec{p}^{\,\mathrm{cm}}_{K}|,
\end{align}
where the upper~(lower) sign represents the positive~(negative) parity resonance, $\mathcal{C}=1$ for the kaon production processes, $|\vec p^{\,cm}_{K}|$ is the 
outgoing kaon momentum measured in the resonance rest frame and is given by, 
\begin{equation}\label{eq:pi_mom}
|\vec{p}^{\,\mathrm{cm}}_{K}| = \frac{\sqrt{(M_R^2-M_{K}^2-M^2)^2 - 4 M_{K}^2 M^2}}{2 M_R}  
\end{equation}
and $E_N$, the outgoing $\Lambda$ baryon energy is
\begin{equation}\label{eq:elam}
  E_N=\frac{M_R^2+M^2-M_{K}^2}{2 M_R}.
\end{equation}

\subsubsection{Spin $\frac{3}{2}$ resonances}
The  general form of the hadronic current for the $s$-channel processes where a positive parity resonance state 
is produced and decays to a lambda~($\Lambda$) and a kaon~($K$) in the final state, is written as~\cite{Athar:2020kqn}:
\begin{eqnarray}
j^\mu\big|_{s}&=& F_{s}^{*} (s) ~\frac{g_{RK\Lambda}}{M_{R}M_{K}} \bar u({p}\,') 
 \epsilon_{\sigma \nu\alpha\beta} p_{R}^{\sigma} \gamma_5 \gamma^{\alpha}p_{K}^{\beta} \left( \frac{{\cal P}^{\nu\delta} (p_{R})}{s-M_{R}^2+ iM_{R} \Gamma_{R}}\right) 
 \Gamma^{\delta\mu}_{\frac32} u({p}\,); \qquad p_R = p+q
\end{eqnarray}
where $\Gamma_{R}$ and $M_{R}$, respectively, are the decay width and mass of the resonance. $F_{s}^{*} (s)$ is defined in Eq.~(\ref{strong_FF_res}), with the value of the cut-off parameter $\Lambda_{R}$ determined for spin $\frac{3}{2}$ resonances to be $\Lambda_{R}=1.62$~GeV. 
$g_{RK\Lambda}$ is the strong coupling strength of the $ R K\Lambda$ vertex, which has been determined using the partial decay width of the resonance to $K\Lambda$ mode. The values of the strong coupling constant of the different resonances are also tabulated in Table~\ref{tab:param-p2}. 

The spin $\frac{3}{2}$ projection operator ${\cal P}_{\alpha \beta} (p_{R})$, in the prescription of Pascalutsa~\cite{Pascalutsa:1999zz}, is given by:
\begin{equation}\label{proj_spin32}
 {\cal P}_{\alpha \beta} (p_{R}) = - (\slashed{p}_{R} + M_{R}) \left(g_{\alpha \beta} - \frac{1}{3} \gamma_{\alpha}\gamma_{\beta} - \frac{1}{3p_{R}^{2}} (\slashed{p}_{R} \gamma_{\alpha} {{p_{R}}}_{\beta} + {{p_{R}}}_{\alpha}\gamma_{\beta} \slashed{p}_{R}) \right).
\end{equation}

The vertex function $\Gamma^{\mu\alpha}_{\frac{3}{2}}$ for the 
positive parity resonances is defined as
\begin{align}\label{eq:vec_32_pos_EM}
  \Gamma^{\mu\alpha}_{\frac{3}{2}} &= {V}^{\mu\alpha}_\frac{3}{2},
  \end{align}
where $V^{\mu}_{\frac{3}{2}}$ represents the vector current, and is  parameterized in terms of the vector couplings $C_{3,4}^{R^{+},R^{0}} $, as
 \begin{align}\label{eq:vectorspinhalf1}
  V^{\mu\alpha}_{\frac{3}{2}} & = \left[\left(\gamma^{\mu}q^{\alpha} - \slashed{q} g^{\alpha\mu} \right) \frac{C_{3}^{R^{+},R^0}}{M} + \left(q \cdot p^{\prime} g^{\alpha\mu} - q^{\alpha} {{p^{\prime}}}^{\mu} \right) \frac{C_{4}^{R^{+},R^0}}{M^2} \right] .
\end{align}
The couplings $C^{R^{+},R^0}_{3,4}$ are derived from the helicity amplitudes extracted from the real photon scattering 
experiments. The explicit relation between the coupling $C_{3,4}^{R^{+},R^0}$ and the helicity amplitude $A_{\frac{1}{2},\frac{3}{2}}^{p,n}$ is given 
by~\cite{Fatima:2022nfn}:
\begin{eqnarray}\label{eq:hel_spin_12}
A_{\frac{1}{2}}^{p,n} &=&  \sqrt{\frac{\pi \alpha}{3M} \frac{M_{R} - M}{M_{R} + M}} \left[\frac{C_{3}^{^{R^{+},R^0}}}{M} (M + M_{R}) + \frac{C_{4}^{^{R^{+},R^0}}}{2M^2} \left(M^2 - M_{R}^{2} \right)  \right], \\
 \label{A32_32}
 A_{\frac{3}{2}}^{p,n} &=&\sqrt{\frac{\pi \alpha}{M} \frac{M_{R} - M}{M_{R} + M}} \left[\frac{C_{3}^{^{R^{+},R^0}}}{M_{R}} \left(M + M_{R} \right) + \frac{C_{4}^{^{R^{+},R^0}}}{2M^2} \left(M_{R}^{2} - M^2 \right)  \right] ,
\end{eqnarray}
where $R^{+}$ and $R^{0}$ correspond, respectively, to the charged and neutral states of the isospin $\frac{1}{2}$ resonances. 
The value of the helicity amplitude $A_{\frac{1}{2}}^{p,n}$ for the different resonances are  quoted in 
Table~\ref{tab:resonance}.


To determine the value of the strong $RK\Lambda$ coupling for spin $\frac{3}{2}$ resonances, we start by writing the general form of 
$RK\Lambda$ Lagrangian given  by Pascalutsa~\cite{Pascalutsa:1999zz}, as:
\begin{equation}
 {\cal L} (x) = \frac{g_{RK\Lambda}}{M_{R}M_{K}} \epsilon_{\mu \nu \alpha \beta} \partial^{\mu} \overline{\Psi}_{R}^{\;\nu} (x) \gamma_{5} \gamma^{\alpha} \psi (x) \partial^{\beta}K (x),
\end{equation}
where $\psi (x)$ is the nucleon field, ${\Psi}_{R} (x)$ 
is the spin $\frac{3}{2}$ resonance field, and $K (x)$ is the kaon field. 

Using the above Lagrangian, one obtains the expression for the decay width in the resonance rest frame as:
\begin{align}\label{eq:12_width}
 \Gamma_{R \rightarrow K\Lambda} &= \frac{\mathcal{C}}{12\pi} \left(\frac{g_{RK\Lambda }}{m_{K}}
 \right)^2  \frac{E_{N} + M}{M_R} |\vec{p}^{\,\mathrm{cm}}_{K}|^3,
\end{align}
where $\mathcal{C}=1$ for kaon production processes, and $|\vec p^{\,\mathrm{cm}}_{K}|$ and $E_N$, respectively, are given in Eqs.~(\ref{eq:pi_mom}) and (\ref{eq:elam}), respectively.

\subsubsection{Kaon resonances}
In the present work, we have considered two kaon resonances in the $t$-channel: a vector meson $K^{*} (892)$ and an axial 
vector meson $K_{1} (1270)$. For details, see Ref.~\cite{Fatima:2020tyh}.
\begin{table*}
  \caption{Properties of kaon resonances included in the present model, with mass $M_R$, spin $J$, isospin 
  $I$, parity $P$, the total decay width $\Gamma$, and the vector 
  $G_{K}^{v}$ and tensor $G_{K}^{t}$ couplings for the kaon resonances. It is to be noted that the couplings $G_{K}^{v}$ 
  and $G_{K}^{t}$ contains both the electromagnetic as well as the strong coupling strengths.}\label{hyperon_resonances}
  \begin{center}
    \begin{tabular*}{160mm}{@{\extracolsep{\fill}}c c c c c c c c}
      \noalign{\vspace{-8pt}}
      \hline \hline
      Resonances               & $M_R$ [GeV] & J\quad   & I \quad    &   P   & $\Gamma$    & $G_{K}^{v}$
      &   $G_{K}^{t}$  \\
      &&&&&(GeV)& &  \\ \hline

      $K^{*}$(892) & $0.89166 \pm 0.00026$ & $1$ &$ \frac{1}{2}$ &$ -$ &   $ 0.0508 \pm 0.0009$  &  $-0.18$ & $0.02$ \\
      \hline
      
      $K_{1}$(1270)  & $1.272 \pm 0.007$ & $1 $ &$ \frac{1}{2}$ &$ +$ &   $ 0.090 \pm 0.020$  & $0.28$ & $-0.28$   \\  \hline
      \hline
    \end{tabular*}
  \end{center}

\end{table*}

The hadronic current for the $K^{*}$ exchange in the $t$-channel is 
obtained as
\begin{eqnarray}\label{had_curr:Kstar}
 J_{\mu} \big|_{K^{*}} &=& ie F_{t}^{\star}(t) \bar{u} (p^{\prime}) \epsilon_{\mu \nu \rho \sigma} q^{\rho} (p^{\prime} - p)^{\sigma} \left(
 \frac{ -g^{\nu \alpha} + (p - p^{\prime})^{\nu} (p - p^{\prime})^{\alpha}/{M_{K^{*}}^{2}}}{t - {M_{K^{*}}^{2}} + i 
 M_{K^{*}} \Gamma_{K^{*}}} \right) \nonumber \\
 &\times&
 \left[G_{K^{*}}^{v} \gamma_{\alpha} + \frac{G_{K^{*}}^{t}}{M + M_{\Lambda}} 
 (\slashed{p}^{\prime} - \slashed{p}) \gamma_{\alpha} \right] u(p),
 \end{eqnarray}
with $G_{K^{*}}^{v} = \kappa_{K K^{*}} g_{K^{*} \Lambda p}^{v}/\mu$ and $G_{K^{*}}^{t} = \kappa_{K K^{*}} g_{K^{*} \Lambda 
p}^{t}/\mu$, where $g_{K^{*} \Lambda 
p}^{v}$ and $g_{K^{*} \Lambda p}^{t}$ are the vector and the tensor couplings, respectively, at the strong $K^{*} \Lambda p$ 
vertex, and $\kappa_{K K^{*}}$ is the coupling strength of the $\gamma K K^{*}$ vertex, $\mu$ is an arbitrary mass factor which is 
introduced to make the Lagrangian dimensionless. $\mu$ is chosen to be 1 GeV. $M_{K^{*}}$ and $\Gamma_{K^{*}}$ are the mass and width of the $K^{*}$ resonance, respectively. Due to the lack of 
the experimental data on the $K^{*}$ and $K_{1}$ resonances, the values of $G_{K^{*}}^{v}$ and $G_{K^{*}}^{t}$ can not be 
determined phenomenologically and are treated as free parameters to be fitted to the experimental data of the $K \Lambda$ 
production and are quoted in Table~\ref{hyperon_resonances}.

The hadronic current for the axial vector kaon $K_{1}$ exchange in the $t$-channel is obtained as
\begin{eqnarray}\label{had_curr:K1}
 J_{\mu} \big|_{K_{1}} &=& ieF_{t}^{\star}(t) \bar{u} (p^{\prime}) [g_{\alpha \mu} q \cdot (p - p^{\prime}) - q_{\alpha} (p - 
 p^{\prime})_{\mu}] \left(\frac{-g^{\alpha \rho} + (p - p^{\prime})^{\alpha} (p - p^{\prime})^{\rho}/{M_{K_{1}}^{2}}}
 {t - {M_{K_{1}}^{2}} + i M_{K_{1}} \Gamma_{K_{1}}} \right) \nonumber \\
 &\times& \left[G_{K_{1}}^{v} \gamma_{\rho} \gamma_{5} + \frac{G_{K_{1}}^{t}}{M + M_{\Lambda}} (\slashed{p}^{\prime} - 
 \slashed{p}) \gamma_{\rho} \gamma_{5} \right] u(p),
 \end{eqnarray}
with $G_{K_{1}}^{v} = \kappa_{K K^{*}} g_{K_{1} \Lambda p}^{v}/\mu$ and $G_{K_{1}}^{t} = \kappa_{K K^{*}} g_{K_{1} \Lambda 
p}^{t}/\mu$, where $\kappa_{K K_{1}}$ is the coupling strength of the electromagnetic $\gamma K K_{1}$ vertex, and $g_{K_{1} \Lambda p}^{v}$ and $g_{K_{1} \Lambda p}^{t}$ 
are the vector and the tensor couplings, respectively, at the strong $K_{1} \Lambda p$ vertex. $M_{K_{1}}$ and $\Gamma_{K_{1}}$ are the mass and width of the $K_{1}$ resonance, respectively. The values of 
$G_{K_{1}}^{v}$ and $G_{K_{1}}^{t}$ are treated as free parameters to be fitted to the experimental data of the $K \Lambda$ 
production and are quoted in Table~\ref{hyperon_resonances}.

In Eqs.~(\ref{had_curr:Kstar}) and (\ref{had_curr:K1}), $F_{t}^{\star}(t)$ is the strong form factor with $\Lambda_{R}$ taken to be the same as  used in the case of Born terms, i.e., $\Lambda_{R} = \Lambda_{B} =$ 0.54 GeV.

\subsubsection{Cross section}

\begin{figure}  
\begin{center}
\includegraphics[width=0.46\textwidth,height=8cm]{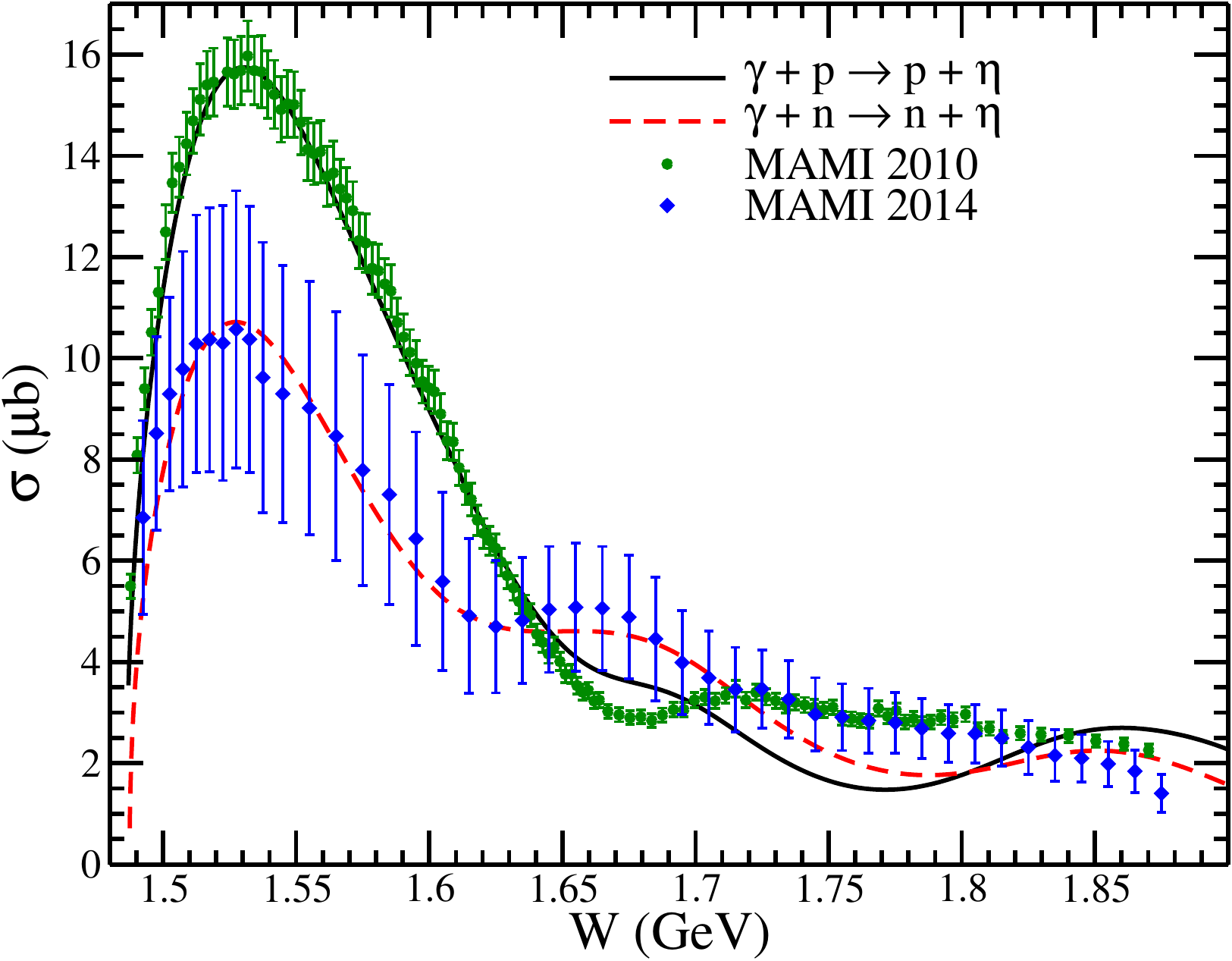}
\includegraphics[width=0.46\textwidth,height=8cm]{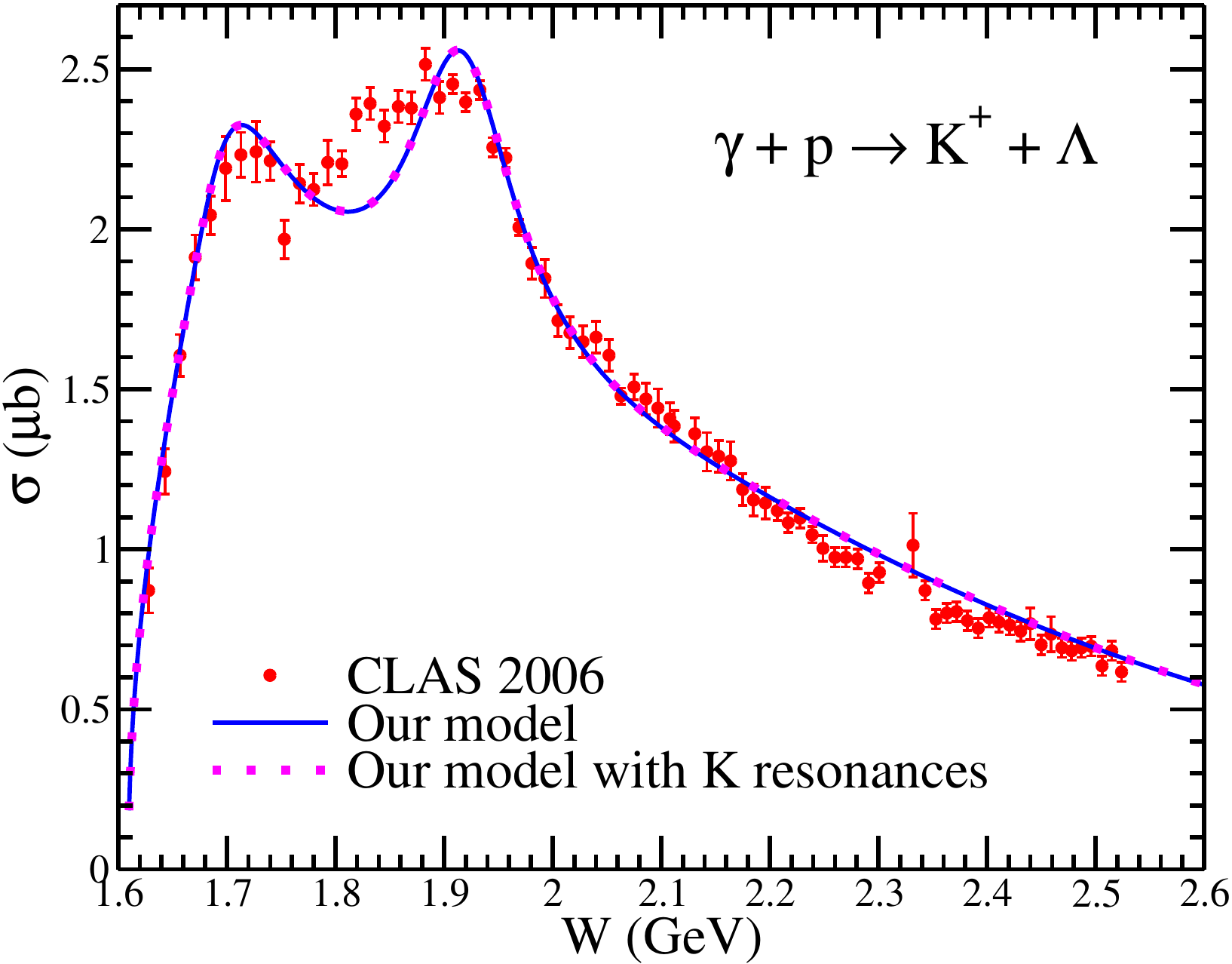}
\caption{(Left panel)~Total cross section $\sigma$ vs. $W$ for $\gamma  p \longrightarrow \eta p$~(solid line) 
and $\gamma  n \longrightarrow \eta n$~(dashed line) processes  using the full model, obtained using the formalism discussed in Ref.~\cite{Fatima:2022nfn, Fatima:2023fez}. The experimental points for the proton target~(solid circle) are 
obtained from MAMI crystal ball collaboration~\cite{CrystalBallatMAMI:2010slt}, and for the neutron target~(solid diamond) we have used the quasifree neutron data from 
MAMI A2 collaboration~\cite{A2:2014pie}. (Right panel)~Comparison of $\sigma$ vs. $W$ for the process $\gamma  p \longrightarrow \Lambda K$ by including all the possible Feynman diagrams depicted in Fig.~\ref{fyn_dia}, without K-meson intermediate state~(dotted line), and with K meson i.e. result of our full calculation~(solid line). Solid circle are the data from the CLAS experiment~\cite{CLAS05}.}
\label{fg_photo_xsec_mami}
\end{center}
\end{figure}


In order to obtain the $\eta$ production as well as the associated particle production cross sections, in the numerical calculations, we consider the following:
\begin{itemize}
\item [(i)] the non-resonant Born terms are obtained using the SU(3) symmetric chiral Lagrangians based on the non-linear sigma model. 

\item [(ii)] for all the spin $\frac{1}{2}$ and $\frac{3}{2}$ resonances, the coupling constant at the strong $R \rightarrow N\eta$ or $R \rightarrow K\Lambda$ vertex is determined using the partial decay width of the resonance into $N\eta$ or $K\Lambda$ decay mode.

\item [(iii)] in the case of $\eta$ production, for the spin $\frac{1}{2}$ resonances, the couplings at the electromagnetic $\gamma NR$ vertex for both the proton and neutron targets, are taken from the PDG~\cite{ParticleDataGroup:2020ssz} and varied within the errors bars to fit the $\eta$ photoproduction data~\cite{Fatima:2022nfn, Fatima:2023fez}. 

\item [(iv)] in the case of $K\Lambda$ production, for the spin $\frac{1}{2}$ resonances, we have used the same values of the electromagnetic couplings as fitted for the $\eta$ photoproduction case, which has been discussed by us in Refs.~\cite{Fatima:2022nfn, Fatima:2023fez}. 

\item [(v)] in the case of $K\Lambda$ production, for the spin $\frac{3}{2}$ resonances, the electromagnetic couplings at $\gamma NR$ vertex for the proton target, are taken from the PDG~\cite{ParticleDataGroup:2020ssz} and varied within the errors bars to fit the $K\Lambda$ photoproduction data~\cite{Fatima:2020tyh}, while for the neutron target we have used the central values of these couplings from the PDG. 
\end{itemize}

In the left panel of Fig.~\ref{fg_photo_xsec_mami}, we have shown the results for $\sigma$ vs. $W$ for $\gamma  p \longrightarrow \eta p$~(solid line) 
and $\gamma  n \longrightarrow \eta n$~(dashed line) processes  using the full model in the region of $W$ from $\eta$ production threshold to $W=1.9$~GeV. We have compared our 
theoretical results with the experimental data obtained by McNicoll et al.~\cite{CrystalBallatMAMI:2010slt} for the MAMI crystal ball
collaboration on the proton target and the quasifree neutron data from Werthmuller et 
al.~\cite{A2:2014pie} for the MAMI A2 collaboration. 
It may be observed from the figure that in the case of $\eta$ production,  our results, with a very few free parameters viz. $\Lambda_B$ and $\Lambda_R$, are in a very good agreement with the 
available experimental data { for the proton in the range of $W \le 1.65$~GeV and for the neutron target in the entire range of $W$ considered in the present work. However, in the case of $\eta$ production from the proton target, our results underestimate the cross sections in the range of $W > 1.65$~GeV, which may be due to the contribution from the higher resonances that need further studies. } 
In the right panel of Fig.~\ref{fg_photo_xsec_mami}, we have presented the results for the total scattering cross section $\sigma$ as a function of center of mass energy $W$ for the photon induced $K\Lambda$ production. The 
theoretical calculations are presented for the full model, which receives contribution from the nonresonant 
Born terms, kaon resonance exchange diagrams, as well as from the $S_{11} (1650)$, $P_{11}(1710)$, $P_{13}(1720)$, $P_{11}(1880)$, 
$S_{11}(1895)$ and $P_{13}(1900)$ resonance excitations.  The results shown by the dotted line also include the contributions from  $K^*$ and $K_1$ resonances, whose contributions are found out to be almost negligible. The theoretical calculations are compared with the experimental data from the CLAS experiment~\cite{CLAS05}. It may be observed from the figure that our results are in a good agreement with the experimental data.

\subsection{Electroproduction}\label{electro}
The electron induced $K\Lambda$ and $\eta$ productions off the nucleon target is given by the reactions
\begin{eqnarray} \label{eq:elecprod}
  e^- (k) + p(p) &\longrightarrow& e^-(k^\prime)  + \Lambda(p^{\prime}) + K^{+} (p_{K}) \,, \\  
  e^- (k) + p(p) &\longrightarrow& e^-(k^\prime)  + p(p^{\prime}) + \eta (p_{\eta}) \,,
\end{eqnarray}
where the four-momentum for each particle is indicated in the parentheses.
The four-momentum of the virtual photon exchanged in the electroproduction is
given by $q = k - k^\prime$. 

The five-fold differential cross section for the electroproduction can be
expressed as~\cite{Donnachie:1978fm, Amaldi:1979vh, Drechsel:1994zx}:
\begin{equation} \label{eq:5fdxs}
 \frac{d\sigma}{d\Omega_l\, dE_l\, d\Omega_{qp_K}}
 = \Gamma\, \frac{d\sigma_\text{v}}{d\Omega_{qp_{K}}} \,,
\end{equation}
with the flux of the virtual photon $\Gamma$ given by
\begin{equation}
 \Gamma = \frac{\alpha}{2 \pi^2}\, \frac{E_l}{E_e}\,
          \frac{K}{Q^2}\, \frac{1}{1-\varepsilon} \,.
\end{equation}
In the above equation, $K =(W^2 - M^2) / 2 M$
denotes the photon equivalent energy, i.e., the laboratory energy necessary
for a real photon to excite a hadronic system with CM energy $W$ and $\varepsilon$ is the transverse polarization parameter of the virtual photon, and is given as
\begin{equation}
 \varepsilon = \left(1 + 2 \frac{|\vec{q}|^2}{\,Q^2}\tan^2\frac{\theta_l}{2}
               \right)^{-1} \,,
\end{equation}
with $Q^2 = - q^2 = -(k-k^\prime)^2$.

The detailed discussion of the electron induced $\eta$ production has been discussed by us in Refs.~\cite{Fatima:2022nfn, Fatima:2023fez}.

The hadronic currents corresponding to the nucleon Born terms for the electroproduction of associated kaons are obtained using the nonlinear sigma model and are  written as: 
\begin{eqnarray}\label{j:s}
J^\mu \arrowvert_{s} &=&ie A_{s}~F_{s}(s) \bar u(p^\prime) \slashed{p}_k \gamma_5 \frac{ \slashed{p} + \slashed{q} + M}
  {s -M^2} {\cal O}_{N}^{\mu} u(p), \\
  \label{j:t}
  J^\mu \arrowvert_{t}&=& ie A_{t}~ F_{t}(t)~F_{K}(Q^2) \bar u(p^\prime)\left[(\slashed{p} - \slashed{p}^{\prime}) \cdot \gamma_{5} \right] 
  u(p) \frac{(2 p_{k}^{\mu} - q^{\mu})}{t - M_{k}^{2}} , \\
  \label{j:ulam}
J^\mu \arrowvert_{u \Lambda} &=&ie A_{u}^{\Lambda} ~F_{u}^{\Lambda} (u) \bar u(p^\prime) {\cal O}_{\Lambda}^{\mu} \frac{ \slashed{p}^{\prime} -\slashed{q} + M_{\Lambda}}
{u - M_{\Lambda}^2} \slashed{p}_k \gamma_5 u(p), \\
\label{j:usig}
J^\mu \arrowvert_{u \Sigma^{0}} &=&ie A_{u}^{\Sigma^{0}} ~F_{u}^{\Sigma^{0}} (u) \bar u(p^\prime) {\cal O}_{\Sigma^0}^{\mu}
\frac{\slashed{p}^{\prime} -\slashed{q} + M_{\Sigma^{0}}} {u -M_{\Sigma^{0}}^2} \slashed{p}_k \gamma_5 u(p), \\
\label{j:CT}
J^\mu \arrowvert_{CT} &=&-i e A_{CT} ~F_{CT} ~F_{C}(Q^2)\bar u(p^\prime) \; \gamma^\mu \gamma_5 \; u(p),
\end{eqnarray}
where the vertex operator ${\cal O}^\mu_{N,\Lambda,\Sigma^0}$ corresponding, respectively, to $\gamma NN$, $\gamma \Lambda\Lambda$, $\gamma\Lambda\Sigma^0$ vertices, is expressed in terms of the $Q^2$ dependent form factors as,
\begin{eqnarray}\label{eq:gNN_vertex}
{\cal O}^\mu_{N,\Lambda,\Sigma^0} &=& F_1^{N,\Lambda,\Sigma^0}(Q^2)\gamma^\mu + F_2^{N,\Lambda,\Sigma^0}(Q^2) i \sigma^{\mu\nu} 
\frac{q_\nu}{2M_{N,\Lambda,\Sigma^0}} .
\end{eqnarray}
The Dirac and Pauli form factors of the nucleon and hyperon viz. $F_{1}^{N,\Lambda,\Sigma^0}(Q^2)$ and $F_{2}^{N,\Lambda,\Sigma^0} (Q^2)$, respectively, are expressed in terms of the 
Sachs' electric~($G_E^{p,n}(Q^2)$) and magnetic~($G_M^{p,n}(Q^2)$) form factors of the nucleons, for which various parameterizations are available in the literature. In the present work, we have taken 
 the parameterization of these form factors from Bradford et al.~\cite{Bradford:2006yz} also known as BBBA05 parameterization. For details, see Ref.~\cite{Fatima:2022nfn}.
 
 The form factors $F_{K}(Q^2)$ and $F_{C} (Q^2)$ are introduced in the $t$-channel and contact term diagrams, respectively, to take into account the hadronic structure at the electromagnetic vertices. The implementation of gauge invariance determines these form factors in terms of the electromagnetic form factors of nucleons and hyperons as:
 \begin{eqnarray}\label{fK}
 F_{K} (Q^2) &=& F_{1}^{p}(Q^2) - F_{1}^{\Lambda} (Q^2) + \sqrt{3} F_{1}^{\Sigma^0} (Q^2) \frac{D-F}{D+3F} \left(\frac{u - M_{\Sigma^0} M_{\Lambda} + MM_{\Sigma^0} - MM_{\Lambda}}{u-M_{\Sigma^0}^2} \right. \nonumber \\
 &+& \left. \left[\frac{M_{\Sigma^0} - M_{\Lambda}}{M+M_{\Lambda}}\right] \left[\frac{u-M^2}{u-M_{\Sigma^0}^2}\right] \right),\\
  \label{fC}
  F_{C} (Q^2) &=&F_{1}^{p}(Q^2) - F_{1}^{\Lambda} (Q^2) + \sqrt{3} F_{1}^{\Sigma^0} (Q^2) \frac{D-F}{D+3F} \left(\frac{u - M_{\Sigma^0} M_{\Lambda} + MM_{\Sigma^0} - MM_{\Lambda}}{u-M_{\Sigma^0}^2}\right).
 \end{eqnarray}      

 The general expression of the hadronic current for the resonance excitation in the $s$-channel is written as,
\begin{eqnarray}\label{jhad:eta_electro}
j^\mu\big|_{s}&=& F_{s}^{*}(s)~ \frac{g_{RK\Lambda}}{f_{K}} \bar u({p}\,') 
 \p_{K} \gamma_5 \Gamma_{s} \left( \frac{\p+\q+M_{R}}{s-M_{R}^2+ iM_{R} \Gamma_{R}}\right) 
 \Gamma^\mu_{\frac12 
 \pm} u({p}\,).
\end{eqnarray}
The vertex function $\Gamma^\mu_{\frac12 \pm}$ for the positive and negative parity resonances is given in Eq.~(\ref{eq:vec_half_pos_EM}), where the vector current $V_{\frac{1}{2}}^{\mu}$ in the case of electroproduction processes is expressed in terms of the form factors $F^{R^+,R^0}_{1,2} (Q^2)$ as:
\begin{eqnarray}\label{eq_eta:nstar_em_vertex}
V_{\frac{1}{2}}^{\mu} &=& \frac{F_1^{R}(Q^2)}{(2 M)^2}(\slashchar{q} q^\mu+Q^2\gamma^\mu) 
+ \frac{F_2^{R}(Q^2)}{2 M} i \sigma^{\mu\nu} q_\nu , \qquad R=R^+,R^0  .
\end{eqnarray} 
The electromagnetic $N-R$ transition form factors  for the charged~($F_{1,2}^{R^+}(Q^2)$)  and neutral~($F_{1,2}^{R^0} (Q^2)$) states are then  related to the helicity amplitudes given by the following relations~\cite{SajjadAthar:2022pjt}: 
\begin{eqnarray}\label{eq2}
 A_{\frac{1}{2}}&=&\sqrt{\frac{2\pi\alpha}{K_R}} 
 \Bra{R,J_Z=\frac{1}{2}}\epsilon_\mu^{+} J_i^\mu \Ket{N,J_Z=\frac{-1}{2}}\zeta \nonumber \\ 
 S_{\frac{1}{2}}&=&-\sqrt{\frac{2\pi\alpha}{K_R}}\frac{|\vec q|}{\sqrt{Q^2}} 
\Bra{R,J_Z=\frac{1}{2}}\epsilon_\mu^{0} J_i^\mu \Ket{N,J_Z=\frac{-1}{2}}\zeta
\end{eqnarray}
where in the resonance rest frame, 
\begin{eqnarray}\label{eq1}
K_R&=&\frac{M_R^{2}-M^2}{2M_R}, \quad \qquad |\vec q |^2=\frac{(M_R^{2}-M^{2}-Q^{2})^2}{4M_R^{2}}+Q^2,\nonumber \\
\epsilon^\mu_{\pm}&=&\mp\frac{1}{\sqrt{2}}(0,1,\pm i,0),\qquad \qquad
\epsilon^\mu_{0}=\frac{1}{\sqrt{Q^{2}}}(|\vec q|,0,0 ,q^0).
\end{eqnarray}
The parameter $ \zeta $ is model dependent which is related to the sign of $R \rightarrow N \pi$, 
and for the present calculation is taken as $\zeta =1$. 

Using Eq.~(\ref{eq1}) in Eq.~(\ref{eq2}), the helicity amplitudes $A_{\frac{1}{2}} (Q^2)$ and $S_{\frac{1}{2}} (Q^2)$ in terms of the electromagnetic
form factors
 $F_1^{R^+,R^0} (Q^2)$ and $F_2^{R^+,R^0} (Q^2)$ are obtained as~\cite{Fatima:2022nfn}:
\begin{eqnarray}\label{eq:hel_em_ff}
A_{\frac{1}{2}}^{p,n} (Q^2)&=& \sqrt{\frac{2 \pi \alpha
}{M}\frac{(M_R \mp M)^2+Q^2}{M_R^2-M^2}} 
\left( \frac{Q^2}{4M^2} F_1^{R^+,R^0}(Q^2) + \frac{M_R \pm M}{2M} F_2^{R^+,R^0}(Q^2) \right) 
\nonumber \\
S_{\frac{1}{2}}^{p,n} (Q^2)&=& \mp \sqrt{\frac{\pi \alpha }{M}\frac{(M_R \pm 
M)^2+Q^2}{M_R^2-M^2}} 
\frac{(M_R \mp M)^2+Q^2}{4 M_R M} \left(  \frac{M_R \pm M}{2M} F_1^{R^+,R^0}(Q^2) - F_2^{R^+,R^0}(Q^2) \right),
\end{eqnarray}
where upper~(lower) sign corresponds to positive~(negative) parity resonances.

The $Q^2$ dependence of the helicity amplitudes~(Eq.~(\ref{eq:hel_em_ff}))  is generally parameterized as~\cite{Tiator:2011pw}:
\begin{equation}\label{eq:ffpar}
{\mathcal A}_{\alpha}(Q^2) = {\mathcal A}_{\alpha}(0) (1+\alpha Q^2)\, e^{-\beta Q^2} ,
\end{equation}
where $ {\mathcal A}_{\alpha}(Q^2)$ are the helicity amplitudes; $A_{\frac12}(Q^2)$ and $S_{\frac12}(Q^2)$ and parameters 
${\mathcal A}_{\alpha}(0)$ are generally determined by a fit to the photoproduction data of the corresponding resonance. 
The values of these parameters for the different nucleon resonances are tabulated in Table~\ref{tab:resonance}.

\subsubsection{Spin $\frac{3}{2}$ resonances}
The general expression of the hadronic current for the resonance excitation in the $s$-channel is written as~\cite{SajjadAthar:2022pjt},
\begin{eqnarray}\label{jhad:eta_electro_32}
j^\mu\big|_{s}&=& F_{s}^{*} (s) ~\frac{g_{RK\Lambda}}{M_{R}M_{K}} \bar u({p}\,') 
 \epsilon^{\sigma \nu\alpha\beta} p_{R_{\sigma}} \gamma_5 \gamma_{\alpha}p_{K_{\beta}} \left( \frac{{\cal P}_{\nu\delta} (p_{R})}{s-M_{R}^2+ iM_{R} \Gamma_{R}}\right) 
 \Gamma^{\delta\mu}_{\frac32} (Q^2) u({p}\,); \qquad p_R = p+q
\end{eqnarray}
The vertex function $\Gamma^{\delta\mu}_{\frac32}$ for the positive parity resonances is given in Eq.~(\ref{eq:vec_32_pos_EM}), where the vector current $V^{\delta\mu}_{\frac{3}{2}}$ in the case of electroproduction processes is expressed in terms of the form factors $C^{R^+,R^0}_{3,4,5} (Q^2)$ as:
\begin{eqnarray}\label{eq_eta:nstar_em_vertex}
V_{\frac{3}{2}}^{\delta\mu} &=&  \left(\gamma^{\mu}q^{\delta} - \slashed{q} g^{\delta\mu} \right) \frac{C_{3}^{R} (Q^2)}{M} + \left(q \cdot p^{\prime} g^{\delta\mu} - q^{\delta} {{p^{\prime}}}^{\mu} \right) \frac{C_{4}^{R} (Q^2)}{M^2} + \left(q^{\delta}q^{\mu} - q^2 g^{\delta\mu} \right) \frac{C_{5}^{R} (Q^2)}{M^2} , \qquad R=R^+,R^0  .
\end{eqnarray} 
The electromagnetic $N-R$ transition form factors for the charged~($C_{3,4,5}^{R^+}(Q^2)$)  and neutral~($C_{3,4,5}^{R^0} (Q^2)$) states are then  related to the helicity amplitudes~($A_{\frac{1}{2}} (Q^2)$, $A_{\frac{3}{2}} (Q^2)$, and $S_{\frac{1}{2}}(Q^2)$), where the explicit relation between the helicity amplitudes $A_{\frac{1}{2}}$, $S_{\frac{1}{2}}$ and the electromagnetic current are given in Eqs.~(\ref{eq2}), and $A_{\frac{3}{2}}$ is expressed in terms of the electromagnetic current by the following expression:
\begin{eqnarray}\label{eq3}
 A_{\frac{3}{2}}&=& \sqrt{\frac{2 \pi \alpha}{K_{R}}} \left<{R, J_{z}^{R} = +\frac{3}{2}} \Big| \epsilon_{\mu}^{+} J^{\mu}_{i} \Big| 
 {N, J_{z}^{N} = +\frac{1}{2}} \right> \zeta.
\end{eqnarray}
The different variables appearing in the above expression are given in Eq.~(\ref{eq1}).

Using Eq.~(\ref{eq1}) in Eqs.~(\ref{eq2}) and (\ref{eq3}), the helicity amplitudes $A_{\frac{1}{2}} (Q^2)$, $A_{\frac{3}{2}} (Q^2)$, and $S_{\frac{1}{2}} (Q^2)$ in terms of the electromagnetic
form factors
 $C_3^{R^+,R^0} (Q^2)$, $C_4^{R^+,R^0} (Q^2)$, and $C_5^{R^+,R^0} (Q^2)$ are obtained as~\cite{Fatima:2022nfn}:
\begin{eqnarray}\label{eq:hel_em_ff_32}
 A_{\frac{1}{2}}(Q^2) &=&  \sqrt{\frac{\pi \alpha}{3M} \frac{\left[(M_{R} - M)^2 + Q^2 \right]}{M_{R}^{2} - M^{2}}} \left[\frac{C_{3}^{R} (Q^2)}{M} (M + M_{R}) + \frac{C_{4}^{R} (Q^2)}{2M^2} \left(M^2 - M_{R}^{2} + Q^2 \right) - C_{5}^{R}(Q^2) \frac{Q^2}{M^2} \right], \\
 \label{A32_32}
 A_{\frac{3}{2}}(Q^2) &=&\sqrt{\frac{\pi \alpha}{M} \frac{\left[(M_{R} - M)^2 + Q^2 \right]}{M_{R}^{2} - M^{2}}} \left[\frac{C_{3}^{R}(Q^2)}{MM_{R}} \left[M(M + M_{R}) + Q^2 \right] + \frac{C_{4}^{R}(Q^2)}{2M^2} \left[M_{R}^{2} - M^2 -Q^2 \right] + C_{5}^{R}(Q^2) \frac{Q^2}{M^2}  \right] , \\
 \label{S12_32}
 S_{\frac{1}{2}}(Q^2) &=& -\sqrt{\frac{\pi\alpha}{6M} \frac{\left[(M_{R} + M)^2 + Q^2 \right]}{M_{R}^{2} - M^{2}} } \left[(M_{R} - M)^2 + Q^2 \right] \left[-\frac{C_{3}^{R}(Q^2)}{MM_{R}} + \frac{C_{4}^{R}(Q^2)}{M^2} + C_{5}^{R}(Q^2) \frac{[M^2 - M_{R}^{2} + Q^2]}{2M^2M_{R}^{2}} \right].
\end{eqnarray}

The parameterization of the $Q^2$ dependence of the helicity amplitudes~(Eq.~(\ref{eq:hel_em_ff_32}))  is given in Eq.~(\ref{eq:ffpar}),
with $ {\mathcal A}_{\alpha}(Q^2)$ being the helicity amplitudes; $A_{\frac12}(Q^2)$, $A_{\frac32} (Q^2)$, and $S_{\frac12}(Q^2)$. The values of the parameters appearing in Eq.~(\ref{eq:ffpar}) for the different spin $\frac{3}{2}$ resonances are tabulated in Table~\ref{tab:resonance}.

\subsubsection{Kaon resonances}\label{Kaon_electro}
In the case of associated particle production induced by the electrons, the expressions for the hadronic current of the kaon exchange diagrams remain the same as given in Eqs.~(\ref{had_curr:Kstar}) and (\ref{had_curr:K1}), except that these terms are multiplied by an additional form factor, which accounts for the electromagnetic structure of the kaons. In literature, generally a monopole form factor of the type 
\begin{equation}
 F_{K^{\star}, K_{1}} (Q^2)  = \frac{1}{\left(1+ \frac{Q^2}{\Lambda_{K^{\star}, K_{1}}} \right)},
\end{equation}
is taken into account with $\Lambda_{K^{\star}} = 0.95$~GeV~\cite{Janssen:2003kk} and $\Lambda_{K_{1}} = 0.55$~GeV~\cite{Janssen:2003kk}.

\subsubsection{Cross section}
\begin{figure}  
\begin{center}
\includegraphics[width=0.46\textwidth,height=8cm]{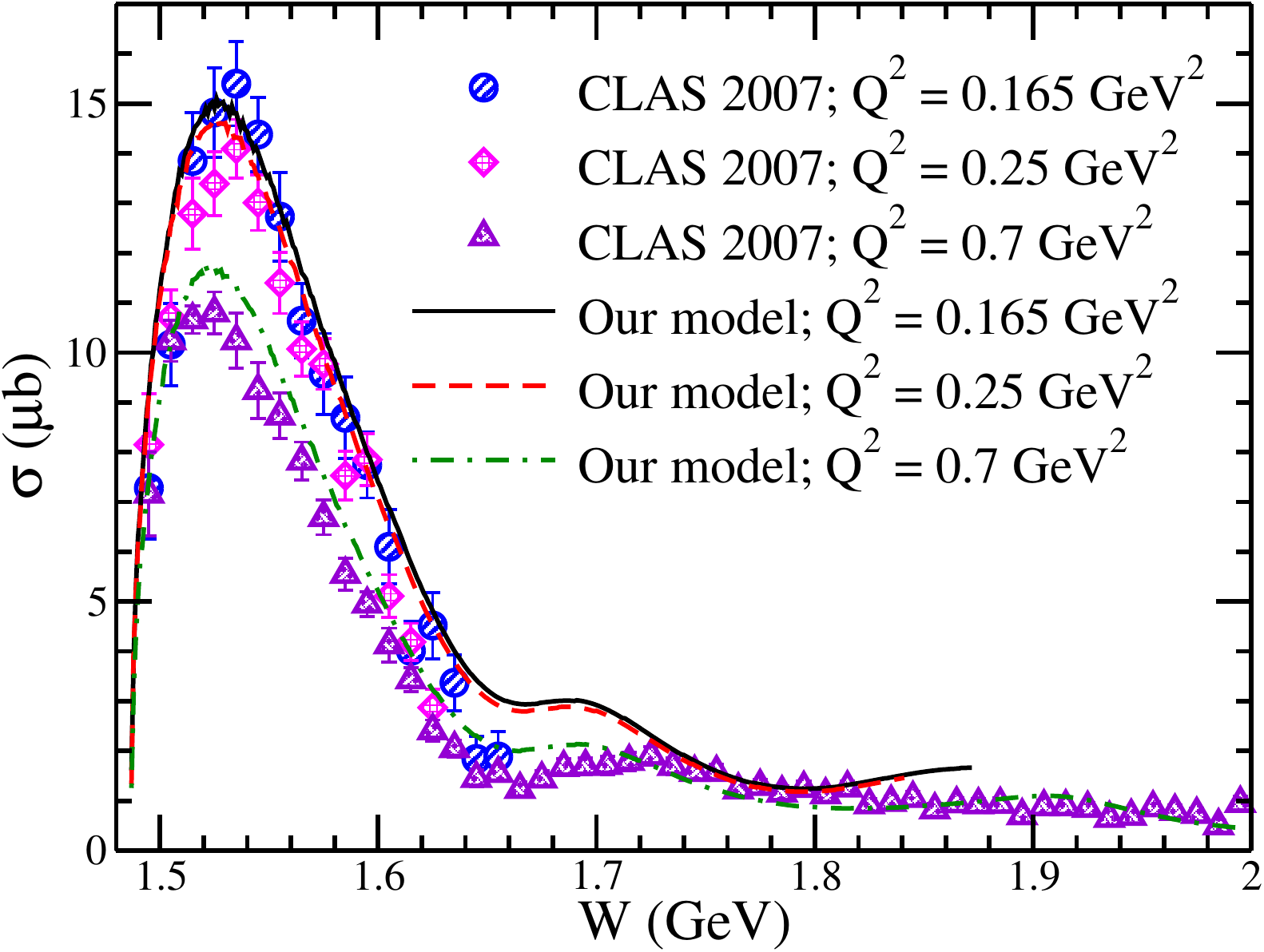}
 \includegraphics[width=0.46\textwidth,height=8cm]{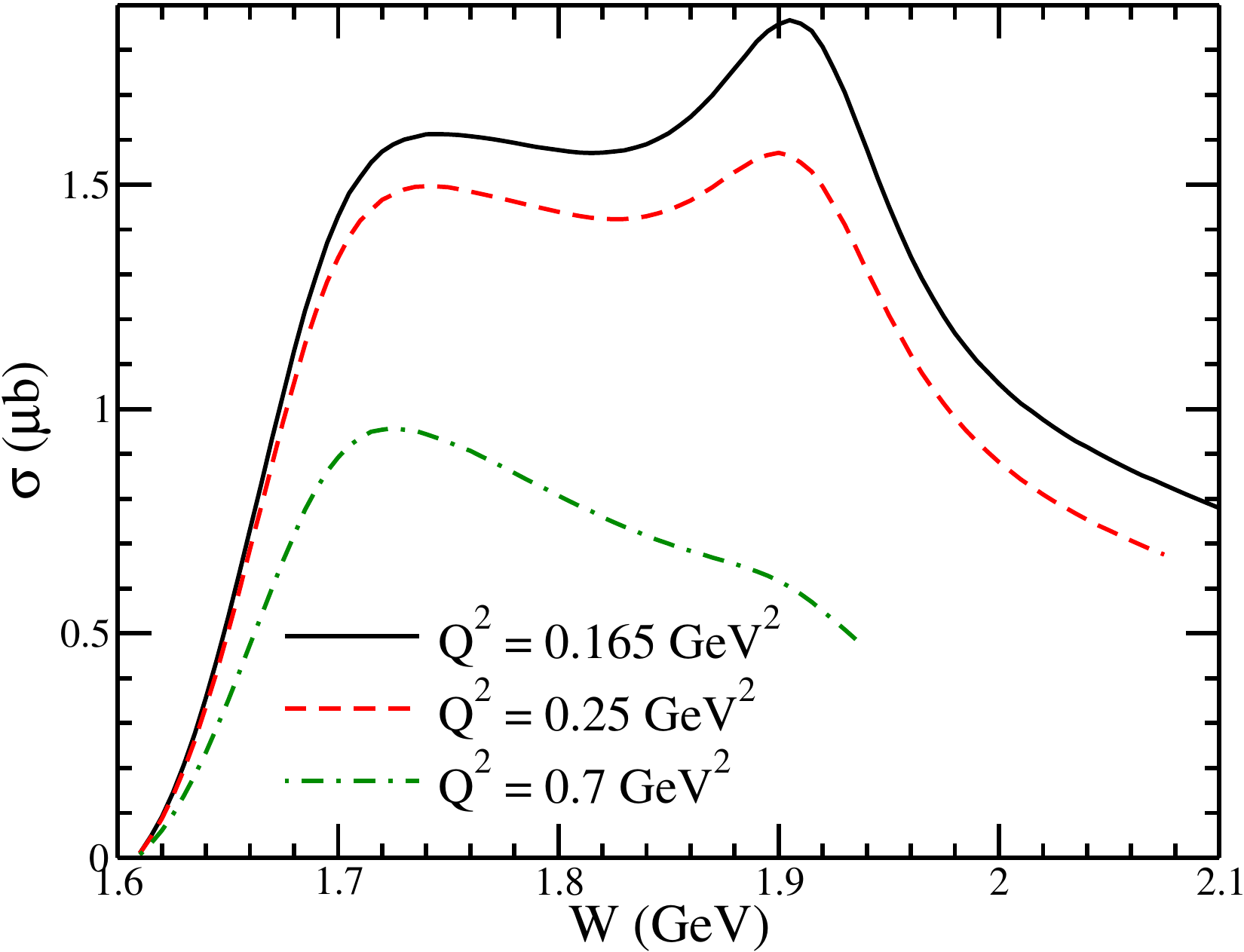}
\caption{(Left panel)~Integrated cross section $\sigma$ vs. $W$ at different $Q^2$ for $\gamma^\ast  p \rightarrow \eta p$ process. The solid line is the result at $Q^2= 0.165$~GeV$^2$, the dashed line is the result at $Q^2= 0.25$~GeV$^2$ and the dash-dotted line is the result at $Q^2= 0.7$~GeV$^2$. These 
 results are presented using the full model which receives contribution from the nonresonant 
Born terms as well as from the nucleon resonance excitations~\cite{Fatima:2023fez}. The experimental 
points are the CLAS 2007 data~\cite{Denizli:2007tq}.  (Right panel)~Integrated cross section $\sigma$ vs. $W$ at different $Q^2$ for $\gamma^\ast  p \rightarrow \Lambda K$ process. The solid line is the result at $Q^2= 0.165$~GeV$^2$, the dashed line is the result at $Q^2= 0.25$~GeV$^2$ and the dash-dotted line is the result at $Q^2= 0.7$~GeV$^2$. These 
 result are presented using the full model which receives contribution from the nonresonant 
Born terms, Kaon resonance exchanges as well as from the nucleon resonance excitations.  }
\label{fg_electro-xsec_mami}
\end{center}
\end{figure}

In order to obtain the electron induced $\eta$ production as well as the $K\Lambda$ production cross sections, in the numerical calculations, we consider the following:
\begin{itemize}
\item [(i)] in the nonresonant Born terms, the nucleon and hyperon vector form factors are expressed in terms of the Sachs' electric and magnetic form factors of the nucleons, for which the BBBA05 parameterization is used.

\item [(ii)] for the spin $\frac{1}{2}$ and $\frac{3}{2}$ resonances, we have used the same values of the coupling constant at the strong $R \rightarrow N\eta$ or $R \rightarrow K\Lambda$ vertex as determined in the case of the photoproduction of $K\Lambda$ or $N\eta$.

\item [(iii)] in the case of $\eta$ production, for the spin $\frac{1}{2}$ resonances, the $Q^2$ dependence of the helicity amplitudes, which are used to determine the electromagnetic $N-R$ transition form factors, is fitted to explain the electroproduction data for the $\eta$ production available from the CLAS experiment~\cite{Fatima:2022nfn, Fatima:2023fez}. 

\item [(iv)] in the case of $K\Lambda$ production, for the spin $\frac{1}{2}$ resonances, we have used the same $Q^2$ dependence of the electromagnetic form factors as fitted for the $\eta$ electroproduction case~\cite{Fatima:2022nfn, Fatima:2023fez}. 

\item [(v)] in the case of $K\Lambda$ production, for the spin $\frac{3}{2}$ resonances, we have fitted the $Q^2$ dependence of the helicity amplitudes, which are used to determine the electromagnetic form factors. 
\end{itemize}

In the left panel of Fig.~\ref{fg_electro-xsec_mami}, we have presented the results for the total cross section 
$\sigma$ for $\gamma^\ast  p \rightarrow \eta p$ process by integrating the angular 
distribution~($\frac{d\sigma}{d\Omega_{qp_{\eta}}}$) given in Eq.~(\ref{eq:5fdxs})
over the polar and 
azimuthal angles as a function of CM 
energy $W$ at different values of $Q^2$ viz. $Q^2= 0.165$~GeV$^{2}$~(solid line), $0.25$~GeV$^{2}$~(dashed line) and $0.7$~GeV$^{2}$~(dash-dotted line). The 
theoretical calculations are presented for the full model, which receives contribution from the nonresonant 
Born terms as well as from the $S_{11}(1535)$, $S_{11} (1650)$, $P_{11}(1710)$, $P_{11}(1880)$ and 
$S_{11}(1895)$ resonance excitations. We have compared our theoretical results with the experimental data 
available from the CLAS 
experiment~\cite{Denizli:2007tq} and found a very good agreement between them at all values of $Q^2$. This has been discussed in Ref.~\cite{Fatima:2023fez} in more detail. In the right panel of Fig.~\ref{fg_electro-xsec_mami}, we have shown the results for the total cross section 
$\sigma$ for $\gamma^\ast  p \rightarrow \Lambda K$ at the three different values of $Q^2$ viz. $Q^2= 0.165$~GeV$^{2}$~(solid line), $0.25$~GeV$^{2}$~(dashed line) and $0.7$~GeV$^{2}$~(dash-dotted line).

\section{Weak production of $K\Lambda$}\label{neutrino}
The weak charged current associated particle production processes induced by the (anti)neutrinos are given by the reactions
\begin{eqnarray}\label{eq:inelastic:reaction}
\nu_{\mu}(k) + n (p) &\longrightarrow& \mu^{-}(k^{\prime}) + \Lambda (p^{\prime}) + K^{+} (p_{K}),\\
\label{eq:inelastic:reaction-2}
{\bar\nu}_{\mu}(k) + p(p) &\longrightarrow& \mu^{+} (k^{\prime}) + \Lambda (p^{\prime}) + K^{0} (p_{K}),
\end{eqnarray}
where the quantities in the parentheses represent the four momentum of the corresponding particles.

\begin{figure}
 \includegraphics[height=5 cm, width=0.9\textwidth]{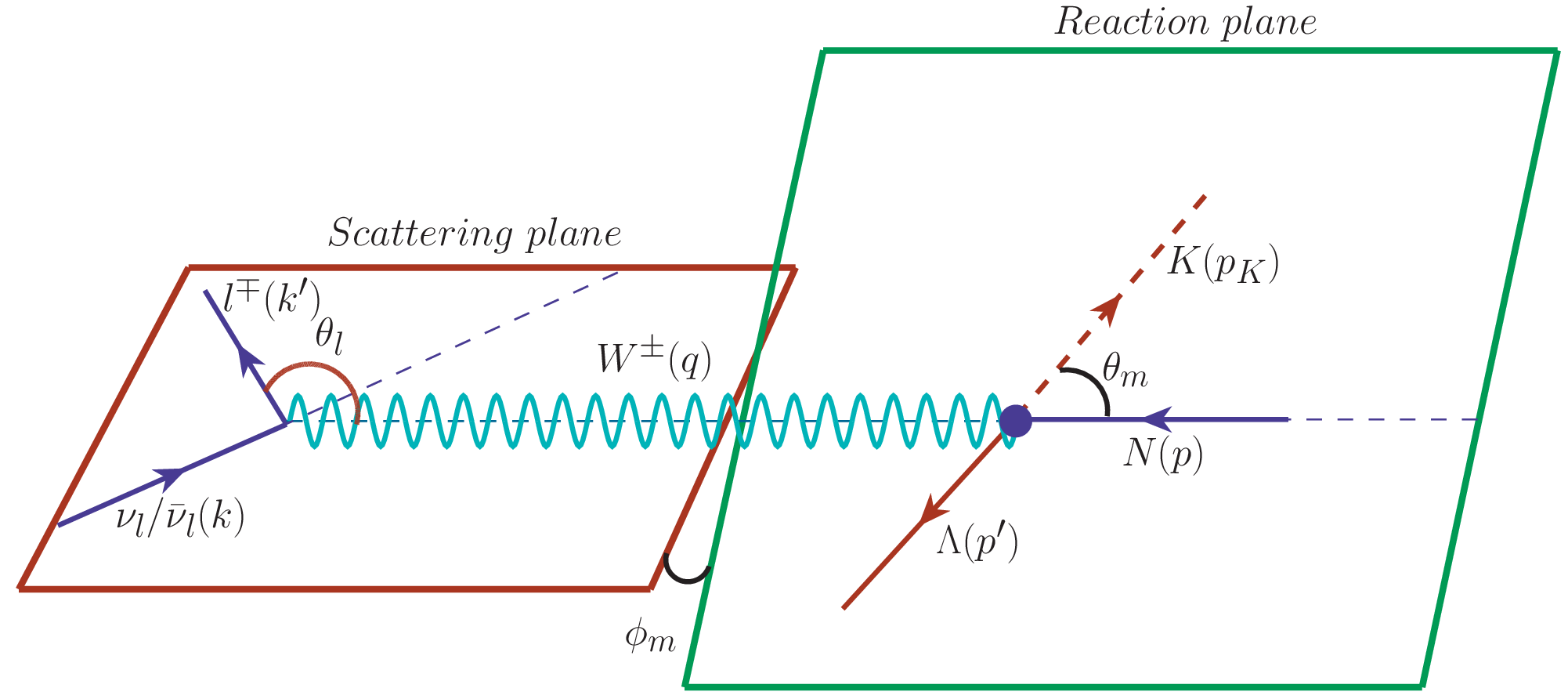}
 \caption{(Anti)neutrino scattering and reaction planes.}
 \label{reactionplane}
\end{figure}

The differential scattering cross section $\frac{d\sigma}{dQ^2}$, for the reactions shown in 
Eqs.~(\ref{eq:inelastic:reaction}) and (\ref{eq:inelastic:reaction-2}) is expressed as
\begin{equation}\label{sigma:weak}
 \frac{d\sigma}{dQ^2} = \int_{W_{min}}^{W_{max}} dW \int_{0}^{2\pi} d\phi_{qp_{K}} \int_{E_{K}^{min}}^{E_{K}^{max}} dE_K \frac{1}{(2\pi)^{4}} \frac{1}{64E_{\nu}^{2}M^2} \frac{W}{|\vec{q}\;|} \overline{\sum} \sum |{\cal M}|^2, 
\end{equation}
where $W_{min}(=M_{\Lambda} + M_K)$ and $W_{max}$ are, respectively, the minimum and maximum CM energies, $E_K^{min}$ and $E_K^{max}$ are, respectively, the minimum and maximum energy of the outgoing kaon and $\phi_{qp_{K}}$ is the azimuthal angle between the scattering plane and the reaction plane as shown in Fig.\ref{reactionplane}.
$\overline{\sum}\sum | \mathcal M |^2  $ is the square of the transition amplitude averaged~(summed) over the 
spins of the initial~(final) states and the transition matrix element is written in terms of the leptonic and the hadronic 
currents as 
\begin{equation}
\label{eq:Gg}
 \mathcal M = \cos\theta_c \frac{G_F}{\sqrt{2}}\, {l_\mu} J^{\mu},
\end{equation}
where $\theta_C(=13.1^o)$ is the Cabibbo angle and $G_F(=1.166 \times 10^{-5}$~GeV$^{-2}$) is the Fermi coupling constant. The leptonic current $l_\mu$ is given by 
\begin{equation}\label{lep_curr}
l_{\mu} = \bar{u} (k^{\prime}) \gamma_{\mu}(1\mp\gamma_5) u (k), 
\end{equation}
where $-(+)$ sign stands for the neutrino~(antineutrino) induced reaction, and $J^{\mu}$ is the hadronic current for $W^{\pm} + N 
\longrightarrow \Lambda + K$  interaction.
The total hadronic current $J^{\mu}$ is written as the sum of the contributions from the nonresonant Born terms~($J^{\mu}_{\text{NR}}$), and the resonance contributions from the spin $\frac{1}{2}$ resonances~($J^{\mu}_{R_{\frac{1}{2}}}$), and spin $\frac{3}{2}$ resonance~($J^{\mu}_{R_{\frac{3}{2}}}$), i.e. 
\begin{equation}\label{allterms}
 J^{\mu}=J^{\mu}_{\text NR} ~+~ J^{\mu}_{R_{\frac{1}{2}}}~+~J^{\mu}_{R_{\frac{3}{2}}}.
\end{equation}

\begin{figure}  
\centering
\includegraphics[height=8cm, width=16cm]{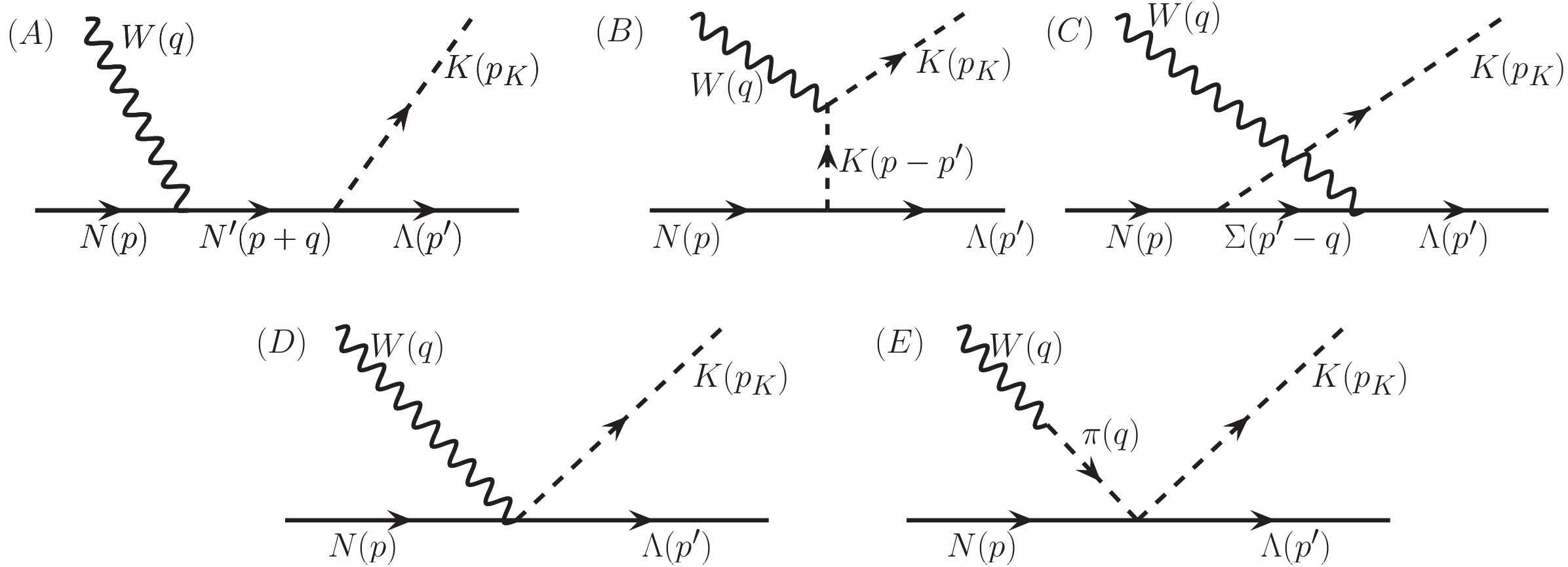}
\caption{Feynman diagrams corresponding to the { nonresonant background terms of the} (anti)neutrino induced $\Delta S=0$ associated particle production processes. (A) $s$-channel, (B) $t$-channel or pion in flight~(PF) term, (C) $u$-channel, (D) contact term~(CT), and (E) pion pole~(PP) term constitute the non-resonant Born diagrams. { Wiggly line represents the intermediate vector meson $W$, solid line represents the baryons like nucleon and hyperons, and the dashed line represents the meson like pion and kaon.}}
\label{Ch12_fig:feyn_app}
\end{figure}

\subsection{Non-resonant contribution}
The nonresonant terms are calculated using the nonlinear sigma model~\cite{Fatima:2020tyh, SajjadAthar:2022pjt}, and the expressions for the  hadronic currents corresponding to the 
diagrams shown in Fig.~\ref{Ch12_fig:feyn_app} are obtained as:
\begin{eqnarray}\label{Ch12_Eqapp:amplitude}
j^\mu \arrowvert_{s} &=& i A_{SY}   \; \bar u (p^\prime) \slashed{p}_K \gamma_5 
	    \frac{{p\hspace{-.5em}/} + {q\hspace{-.5em}/} + M}{(p+q)^2-M^2} 
	    \left[V^{\mu} - A^{\mu} \right] u(p) \nonumber \\
j^\mu \arrowvert_{PF} &=& i A_{TY} \; f_{PF}(Q^2)\; (M+M_\Lambda) \; \bar u (p^\prime) \gamma_5 \; u (p) \;\;
	    \frac{2 p_K^\mu - q^\mu }{(p-p^\prime)^2-m_K^2}\nonumber \\
j^\mu \arrowvert_{u} &=& i A_{UY} \; \bar u (p^\prime) \left[V^{\mu} - A^{\mu} \right]
	      \frac{{p\hspace{-.5em}/} - \slashed{p}_K + M_{\Sigma}}{(p - p_K)^2-M_{\Sigma}^2}  
	      \slashed{p}_K \gamma_5 u (p)\nonumber\\
j^\mu \arrowvert_{CT} &=& i A_{CT} \; \bar u (p^\prime) 
		    \left[ \gamma^\mu f_{\rho}((q-p_{K})^2) + B_{CT} \; f_{CT}(Q^2)\;\gamma^\mu  \gamma_5 \right] u (p) \nonumber\\
j^\mu \arrowvert_{PP} &=& i A_{\pi} \; f_{\rho}((q-p_{K})^2) \; \bar u (p^\prime) 
	      \left[{q\hspace{-.5em}/} + \slashed{p}_K\right] u(p) \frac{q^\mu}{q^2-m_\pi^2}
\end{eqnarray}
where,

\begin{table}
\begin{center}
\centering
\renewcommand{\arraystretch}{1.2}
\begin{tabular*}{135mm}{@{\extracolsep{\fill}}ccccccc} \hline \hline
 $A_{CT}$	            &$B_{CT}$		      &     $A_{SY}$
& \multicolumn{2}{c}{$A_{UY}$}                        & $A_{TY}$                     &  $A_{\pi }$ \\ 
                                                                       &                         &    
                                               & $Y^\prime=\Sigma$             & $Y^\prime=\Lambda$  &       
                                               &             \\     \hline
$-\frac{\sqrt{3}}{2f_{K}}$   & $\frac{-1}{3}(D+3F)$ & $\frac{D+3F}{2\sqrt{3}f_{K}}$ &  $\frac{\sqrt{2}}{3f_{K}}(D-F)$ & 0 & 
$\frac{D+3F}{2\sqrt{3}f_{K}}$ & -$\frac{\sqrt{3}}{4f_{K}}$ \\ \hline\hline
\end{tabular*}
\caption{Constant factors  appearing in the hadronic current in Eq.~(\ref{Ch12_Eqapp:amplitude}). }
\label{Ch12_tb_app:currents}
\end{center}
\end{table}
\begin{eqnarray}
 V^\mu &=& f_1^{BB^{\prime}} (Q^{2}) \gamma^\mu + i \frac{f_2^{BB^{\prime}}(Q^{2})}{M+M^{\prime}} \sigma^{\mu \nu} q_\nu \\
 A^{\mu} &=& g_1^{BB^{\prime}}(Q^{2}) \gamma^\mu \gamma_{5} + g_3^{BB^{\prime}}(Q^{2})\frac{2 q^\mu}{M+M^{\prime}}\gamma_5
\end{eqnarray}
are the vector~($V^{\mu}$) and axial-vector~($A^{\mu}$) transition currents for $BB^{\prime} = NN^\prime$ for the $s$-channel diagram, and $BB^{\prime} = YY^\prime$ for the $u$-channel diagram, with $N,N^\prime = n,p$ and $Y,Y^{\prime}=\Lambda, \Sigma^{\pm}$. 
The vector and axial-vector form factors $f_{1,2}^{BB^{\prime}} (Q^2)$ and $g_{1,
3}^{BB^{\prime}}(Q^2)$ are determined assuming the Cabibbo theory and the various symmetry properties of the weak hadronic 
current~(for details see Ref.~\cite{SajjadAthar:2022pjt}).  The isovector vector form factors 
$f_{1,2}^{BB^\prime}(Q^2)$ are expressed in terms of the Dirac~($F_1^{p,n} (Q^2)$) and Pauli~($F_2^{p,n} (Q^2)$) form factors, discussed in Section~\ref{electro}, for the 
proton and the neutron,  using the relationships
\begin{eqnarray}\label{Eq_eta:f1v_f2v}
f_{1,2}^{NN^\prime}(Q^2)&=&F_{1,2}^p(Q^2)- F_{1,2}^n(Q^2), \\
f_{1,2}^{YY^\prime}(Q^2)&=&- \sqrt{\frac{3}{2}}F_{1,2}^n(Q^2).
\end{eqnarray}
These electromagnetic form factors may be rewritten in terms of the electric~($G_{E}^{N} (Q^2)$) and magnetic~($G_{M}^{N} (Q^2)$) Sachs' form factors.
 
The axial-vector form factor $g_1^{BB^\prime}(Q^2)$ is parameterized as
\begin{equation}\label{Eq_eta:fa}
g_1^{BB^\prime}(Q^2)=g_1^{BB^\prime}(0)~\left[1+\frac{Q^2}{M_A^2}\right]^{-2},
\end{equation}
where $g_1^{NN^\prime}(0)=D+F=1.267$ and $g_1^{YY^\prime}(0)=-\sqrt{\frac{3}{2}}D$ is the axial-vector charge for the $N-N^\prime$ and $Y-Y^\prime$ transitions, respectively, and $M_A$ is the axial dipole mass, which in the numerical calculations is taken as the world average value i.e. $M_A = 1.026$~GeV~\cite{Bernard:2001rs}.

On the other hand, the pseudoscalar form factor $g_3^{BB^\prime}(Q^2)$ is expressed  in terms of $g_1^{BB^{\prime}}(Q^2)$ 
using the PCAC hypothesis and Goldberger-Treiman relation as
\begin{equation}\label{Eq:fp_nucleon}
g_3^{BB^\prime}(Q^2)=\frac{(M+M^\prime)^2g_1^{BB^\prime}(Q^2)}{2(m_\pi^2+Q^2)},
\end{equation}
with $m_\pi$ being the pion mass.

The form factors $f_{CT}(Q^2)$, $f_{PF}(Q^2)$ and 
$f_{\rho}((q-p_{K})^2)$ are introduced in the contact, pion pole and pion in flight terms to take into account the hadronic 
structure. It may be observed from the Feynman diagrams~(Fig.~\ref{Ch12_fig:feyn_app}) that the pion in flight term is purely 
vector in nature while the pion pole diagram is possible only with axial-vector current. In the case of contact term, the 
term associated with $B_{CT}$ represents the vector part of the weak hadronic current while the term with $\gamma^{\mu}$ is 
associated with the axial-vector part. CVC hypothesis imposes the following condition on the form factors $f_{CT}(Q^2)$ and 
$f_{PF}(Q^2)$, {  i.e.},
\begin{eqnarray}
 f_{CT} (Q^2) &=& f_{1}^{NN^\prime} (Q^2) - 2F_{1}^{n} (Q^2) \; \left(\frac{D-F}{D+3F}\right) \; \left(\frac{u - M_{\Sigma} 
 M_{\Lambda} + MM_{\Sigma} - MM_{\Lambda}}{M_{\Sigma}^{2} - u} \right),\\
 f_{PF} (Q^2) &=& 2F_{1}^{n} (Q^2) \left(\frac{D-F}{D+3F}\right) \; \left(\frac{(M+M_{\Sigma})(u-M_{\Lambda}^2)}{(M + 
 M_{\Lambda})(M_{\Sigma}^2 - u)}\right) - f_{1}^{NN^\prime} (Q^2),
\end{eqnarray}
where $u=(p-p_{K})^2$, $f_{1}^{NN^\prime} (Q^2) = F_{1}^{p} (Q^2) - F_{1}^{n} (Q^2)$ is the vector form factor with $F_{1}^{p} 
(Q^2)$ and $F_{1}^{n} (Q^2)$ being the nucleon electromagnetic form factors. The form factor $f_{\rho} (Q^2)$ corresponding to the axial-vector current is given by~\cite{SajjadAthar:2022pjt, Hernandez:2007qq}:
\begin{equation}
 f_{\rho}(Q^2) = \frac{1}{1+Q^2/m_{\rho}^2}; \qquad \qquad {\rm with } \; m_\rho = 0.776~\text{ GeV}.
\end{equation}

\subsection{Resonance contribution}
\subsubsection{Spin $\frac{1}{2}$ resonance excitations}
Next, we discuss the positive and negative parity resonance excitation mechanism for the weak interaction induced $K\Lambda$ production. 
The general expression of the hadronic current for the $s$-channel resonance excitation and their subsequent decay to $K\Lambda$ mode are given in Eq.~(\ref{jhad:eta_electro}), where the vertex factor $\Gamma_{\frac{1}{2}\pm}^{\mu}$ is now written as
\begin{align}\label{eq:vec_half_pos}
  \Gamma^{\mu}_{\frac{1}{2}^+} &= {V}^{\mu}_\frac{1}{2} - {A}^{\mu}_\frac{1}{2},
  \end{align}
  for the positive parity resonance, and as
\begin{align}\label{eq:vec_half_neg}
  \Gamma^{\mu}_{\frac{1}{2}^-} &= \left({V}^{\mu}_\frac{1}{2} - {A}^{\mu}_\frac{1}{2} \right) \gamma_5 ,
  \end{align}
  for the negative parity resonance. The vector and axial-vector vertex factors for the weak charged current interaction processes are given by
\begin{align}  \label{eq:vectorspinhalfcurrent}
  V^{\mu}_{\frac{1}{2}} & =\frac{{f_{1}^{CC}}(Q^2)}{(2 M)^2}
  \left( Q^2 \gamma^\mu + {q\hspace{-.5em}/} q^\mu \right) + \frac{f_2^{CC}(Q^2)}{2 M} 
  i \sigma^{\mu\alpha} q_\alpha ,  \\
    \label{eq:axialspinhalfcurrent}
  A^{\mu}_\frac{1}{2} &=  \left[{g_1^{CC}}(Q^2) \gamma^\mu  +  \frac{g_3^{CC}(Q^2)}{M} q^\mu\right] \gamma_5 ,
\end{align}
where $f_i^{CC}(Q^2)$~($i=1,2$) are the isovector $N-R$ transition form factors which, in turn, are expressed in terms of the 
charged~($F_{i}^{R+} (Q^2)$) and neutral~($F_{i}^{R0} (Q^2)$) electromagnetic $N-R$ transition form factors as:
\begin{equation}\label{eq:f12vec_res_12}
f_i^{CC}(Q^2) = F_i^{R+}(Q^2) - F_i^{R0}(Q^2), \quad \quad i=1,2
\end{equation}
Further, these form factors are related to the helicity amplitudes as discussed in Section~\ref{electro}.

The axial-vector current consists of two form factors viz. $g_1^{CC}(Q^2)$ and $g_3^{CC}(Q^2)$, which are determined 
assuming the PCAC hypothesis and the pion pole dominance 
of the divergence of the axial-vector current through the generalized GT relation for $N 
- R$ transition~\cite{SajjadAthar:2022pjt}. 

The axial-vector coupling $g_{1}^{CC}$ at $Q^2=0$ is obtained as~\cite{Fatima:2022nfn}:
\begin{equation}\label{eq:g1_pos}
g_1^{CC}(0)= 2 g_{RN\pi},
\end{equation}
with $g_{RN\pi}$ being the coupling strength for $R \to N\pi$ decay, 
which has been determined by the partial decay width of the resonance and tabulated in Table~\ref{tab:param-p2}. Since no information about the $Q^2$ dependence of 
the axial-vector form factor for the $N-R$ transition is known experimentally, therefore, a dipole form is assumed:
\begin{equation}\label{ga:CC}
 g_1^{CC}(Q^2) = \frac{g_1^{CC}(0)}{\left(1+\frac{Q^2}{M_{A}^2}\right)^2},
\end{equation}
with $M_{A}=1.026$~GeV, and the pseudoscalar form factor $g_3^{CC}(Q^2)$  is given by
\begin{equation}\label{eq:fp_res_spinhalf}
g_{3}^{CC}(Q^2) = \frac{(MM_{R}\pm M^{2})}{m_{\pi}^{2}+Q^{2}} g_1^{CC}(Q^2) ,
\end{equation}
where $+(-)$ sign is for positive~(negative) parity resonances. However, the contribution of $g_{3}^{CC} (Q^2)$ being directly proportional to the lepton mass is almost negligible for the charged current $\nu_{\mu}(\bar{\nu}_{\mu})$ induced processes.

\subsubsection{Spin $\frac{3}{2}$ resonance excitations}
In the case of positive parity spin $\frac{3}{2}$ resonance excitation and its subsequent decay to $K\Lambda$ channel, the expression of the hadronic current for the $s$-channel diagram is given in Eq.~(\ref{jhad:eta_electro_32}), where the vertex factor $\Gamma_{\frac{3}{2}}^{\mu\alpha}$ is now written as
\begin{align}\label{eq:vec_half_pos}
  \Gamma^{\mu\alpha}_{\frac{3}{2}} &= {V}^{\mu\alpha}_\frac{3}{2} - {A}^{\mu\alpha}_\frac{3}{2}.
  \end{align}
 The vector and axial-vector vertex factors for the weak charged current interaction processes are given by
\begin{eqnarray}  \label{eq:vectorspinhalfcurrent}
    V^{ \alpha\mu}_{\frac{3}{2}} &=& \left(\gamma^{\mu}q^{\alpha} - \slashed{q} g^{\alpha\mu} \right) \frac{C_{3}^{V} (Q^2)}{M} + \left(q \cdot p^{\prime} g^{\alpha\mu} - q^{\alpha} {{p^{\prime}}}^{\mu} \right) \frac{C_{4}^{V} (Q^2)}{M^2} + \left(q^{\alpha}q^{\mu} - q^2 g^{\alpha\mu} \right) \frac{C_{5}^{V} (Q^2)}{M^2},  \\
  \label{eq:axial_3half_pos}
  A^{\alpha\mu}_{\frac{3}{2}} &=&- \left[ \frac{{ C}_3^A (Q^2)}{M} (g^{\mu \alpha} {q\hspace{-.5em}/} \, - q^{\alpha} \gamma^{\mu})+
  \frac{{ C}_4^A (Q^2)}{M^2} (g^{\mu \alpha} q\cdot p' - q^{\alpha} p'^{\mu})+
 {{ C}_5^A (Q^2)} g^{\mu \alpha}
  + \frac{{ C}_6^A (Q^2)}{M^2} q^{\alpha} q^{\mu}\right] \gamma_5 ,
\end{eqnarray}
where ${ C}^V_i (Q^2)$ and ${ C}^A_i (Q^2)$ are the vector and axial-vector form factors. 

The isovector ${C}_i^V (Q^2); (i=3,4,5)$ form factors  are written in terms of the electromagnetic $C^{R+}_i(Q^2)$ and 
$C^{R0}_i(Q^2)$ form factors through a simple relation~\cite{Athar:2020kqn} as
\begin{equation}\label{eq:civ_NC}
{ C}_i^V (Q^2)= C^{R+}_i (Q^2) - C^{R0}_i (Q^2) ; \;\; \qquad i = 3,4,5\, .
\end{equation}
 Further, the electromagnetic form factors are related to the helicity amplitudes as discussed in Section~\ref{electro}.

The form factors ${C}_i^A(Q^2), \; (i=3,4,5,6)$ corresponding to the axial current have not been studied in the case of 
higher resonances. The earlier calculations have used PCAC to determine ${C}_5^A(Q^2)$ and ${C}_6^A(Q^2)$ and taken other 
form factors to be zero~\cite{Adler:1968tw}. In view of this, we have also taken a simple model for the determination of the axial form factors 
based on PCAC and GT relation and write ${C}_6^A(Q^2)$ in terms of ${C}_5^A(Q^2)$ as
\begin{equation}\label{c6_CC}
 C_6^A(Q^2) = C_5^A(Q^2) \frac{M^2}{Q^2 + m_\pi^2} .
\end{equation}
For ${C}_5^A(Q^2)$, a dipole form has been assumed 
 \begin{equation}\label{c5a-r}
{C}_5^A(Q^2) = \frac{{C}_5^A(0)}{ \left( 1 + Q^2 /{M_A^{\it R}}^2 \right)^2 } ,
\end{equation}
with ${C}_5^A(0)= -2 g_{R N \pi}$~\cite{SajjadAthar:2022pjt} for the isospin $\frac{1}{2}$ resonances considered in the present work, with
$g_{R N \pi}$ being the coupling for $R \longrightarrow N \pi$ decay for each resonance $R$. $M_{A}^{\it R}$ is taken as 
$1.026~{\rm GeV}$. ${C}_3^A(Q^2)$ as well as ${C}_4^A(Q^2)$ are taken as zero. 

\subsubsection{Spin 1 Kaon resonances}
We have taken the same form of the hadronic current for the spin 1 kaon resonances being exchanged in the $t$-channel as that in the case of electroproduction, discussed in Sec.~\ref{Kaon_electro}.

\section{Result and discussion}\label{results}

 \begin{figure}  
\begin{center}
\includegraphics[width=0.48\textwidth,height=8cm]{neutrino_associated_individual_terms.eps}
\includegraphics[width=0.48\textwidth,height=8cm]{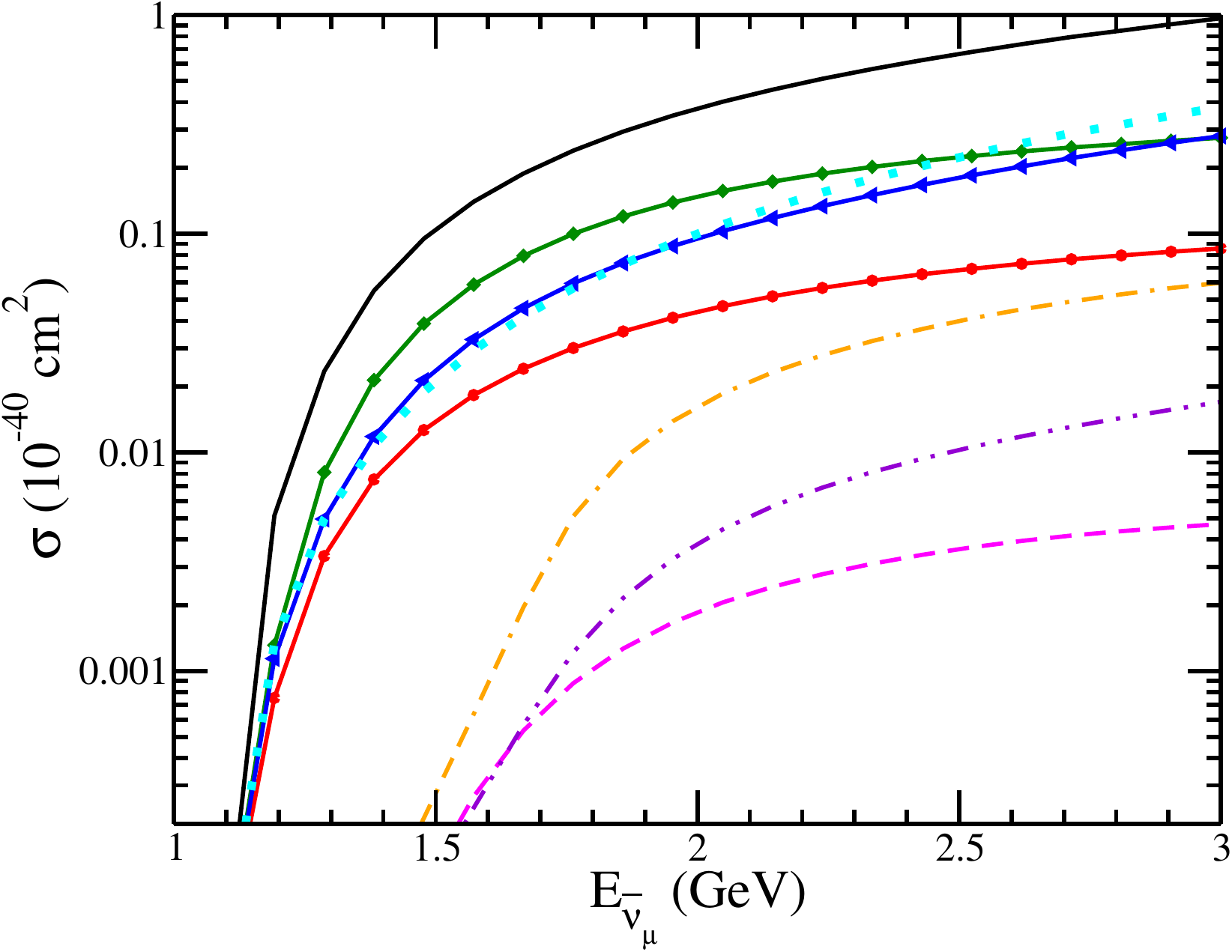}
\caption{Cross section for $\nu_\mu + n \rightarrow \mu^- + \Lambda + K^+$~(left panel) and $\bar\nu_\mu + p \rightarrow \mu^+ + \Lambda + K^0$~(right panel) processes. Solid~(dotted) line represents the results of the full model~(only nonresonant Born terms), solid line with circle, diamond and left triangle, respectively, represents the individual contribution from $S_{11}(1650)$, $P_{11}(1710)$, and $P_{13}(1720)$ resonances. Dashed line, double-dash-dotted line, and double-dotted-dashed line, respectively, 
represents the individual contribution from $P_{11}(1880)$, $S_{11}(1895)$, and $P_{13}(1900)$ resonances.}
\label{assoc-neut}
\end{center}
\end{figure}

 In this section, we present the numerical results for the (anti)neutrino induced $K\Lambda$ production from the nucleon target, using the following inputs for evaluating the contributions from the various terms:
\begin{itemize}
\item [(i)] in the case of Born terms
\begin{itemize}
 \item [(a)] the nucleon and hyperon weak vector form factors are expressed in terms of the electromagnetic form factors of the nucleons, which in turn are expressed in terms of the Sachs' electric and magnetic form factors of the nucleons, for which the BBBA05 parameterization is used.
 
 \item [(b)] the nucleon and hyperon axial vector form factors are obtained assuming the PCAC hypothesis and the generalized Goldberger-Treiman relation.   
\end{itemize}

\item [(ii)] in the spin $\frac{1}{2}$ and $\frac{3}{2}$ resonances
\begin{itemize}
 \item [(a)] the value of the strong coupling constant at the $R \rightarrow K\Lambda$ vertex, determined from the photoproduction of $K\Lambda$ is used.
 
 \item [(b)] the weak vector $N-R$ transition form factors are expressed in terms of the electromagnetic $N-R$ transition form factors, as discussed in Sec.~\ref{neutrino}.

 \item [(c)] for the determination of the axial vector $N-R$ transition form factors, the PCAC hypothesis along with the generalized Goldberger-Treiman relation is used. For the $Q^2$ dependence of the  axial vector $N-R$ transition form factor, a dipole form factor is used with $M_{A} = 1.026$~GeV. 
\end{itemize}
\end{itemize}

 \begin{figure}
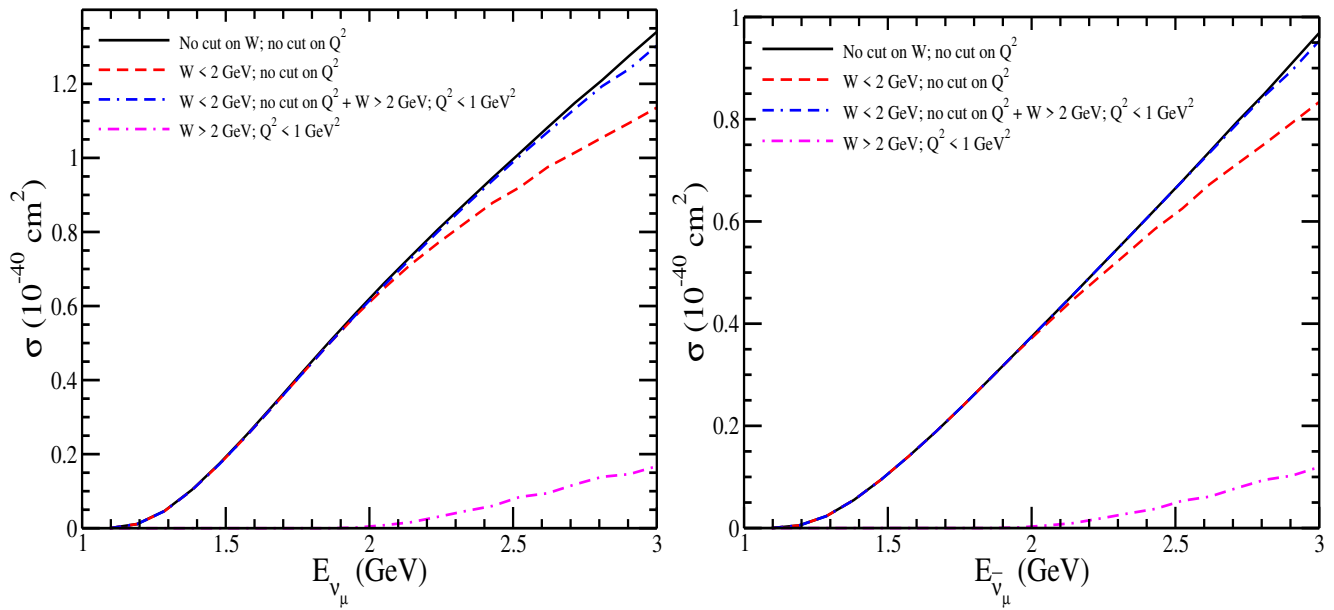
  
\begin{center}
\includegraphics[width=0.48\textwidth,height=8cm]{sigma_nu_associated.eps}
\includegraphics[width=0.48\textwidth,height=8cm]{sigma_anu_associated.eps}
\caption{Cross section for $\nu_\mu + n \rightarrow \mu^- + \Lambda + K^+$~(left panel) and $\bar\nu_\mu + p \rightarrow \mu^+ + \Lambda + K^0$~(right panel) processes. Solid line represents the results of the full model when no cut on $W$ and $Q^2$ are included in the numerical calculations. Dashed line represents the results of the full model when a cut of $W<2$~GeV and no cut on $Q^2$ is applied. Dash-dotted line represents the results of the full model when a cut of $W>2$~GeV and $Q^2< 1$~GeV$^{2}$ is applied. The double-dash-dotted line represents the results of the full model when a cut of $W<2$~GeV and no cut on $Q^2$ is applied along with the contribution of the cross section when $W>2$~GeV and $Q^2<1$~GeV$^{2}$.}
\label{assoc-neut-cut}
\end{center}
\end{figure}

 \begin{figure}
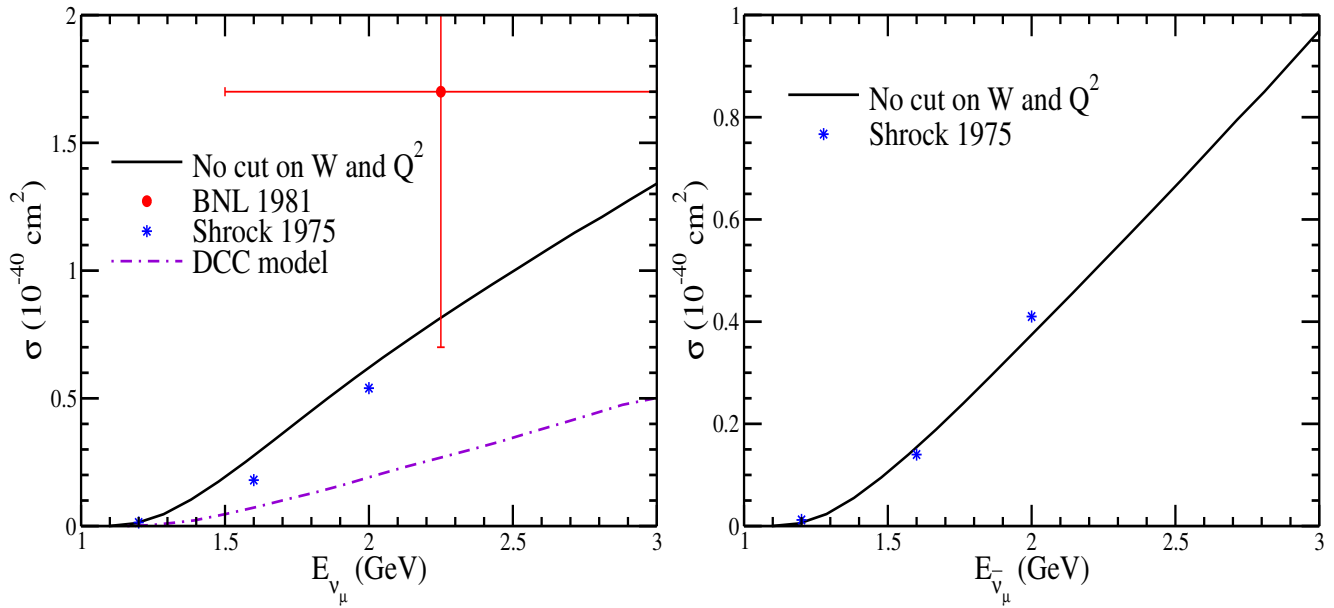
  
\begin{center}
\includegraphics[width=0.48\textwidth,height=8cm]{sigma_nu_associated_compare.eps}
\includegraphics[width=0.48\textwidth,height=8cm]{sigma_anu_associated_compare.eps}
\caption{Cross section for $\nu_\mu + n \rightarrow \mu^- + \Lambda + K^+$~(left panel) and $\bar\nu_\mu + p \rightarrow \mu^+ + \Lambda + K^0$~(right panel) processes. Solid line represents the results of the full model when no cut on $W$ and $Q^2$ are included in the numerical calculations. Dash-dotted line represents the results of the DCC model~\cite{Nakamura:2015rta}, while the star represents the results tabulated in Table-II of Shrock~\cite{Shrock:1975an}. Solid circle is the experimental results of the BNL experiment~\cite{Baker:1981tx}.}
\label{assoc-neut-compare}
\end{center}
\end{figure}

\begin{figure}  
\begin{center}
\includegraphics[width=0.48\textwidth,height=8cm]{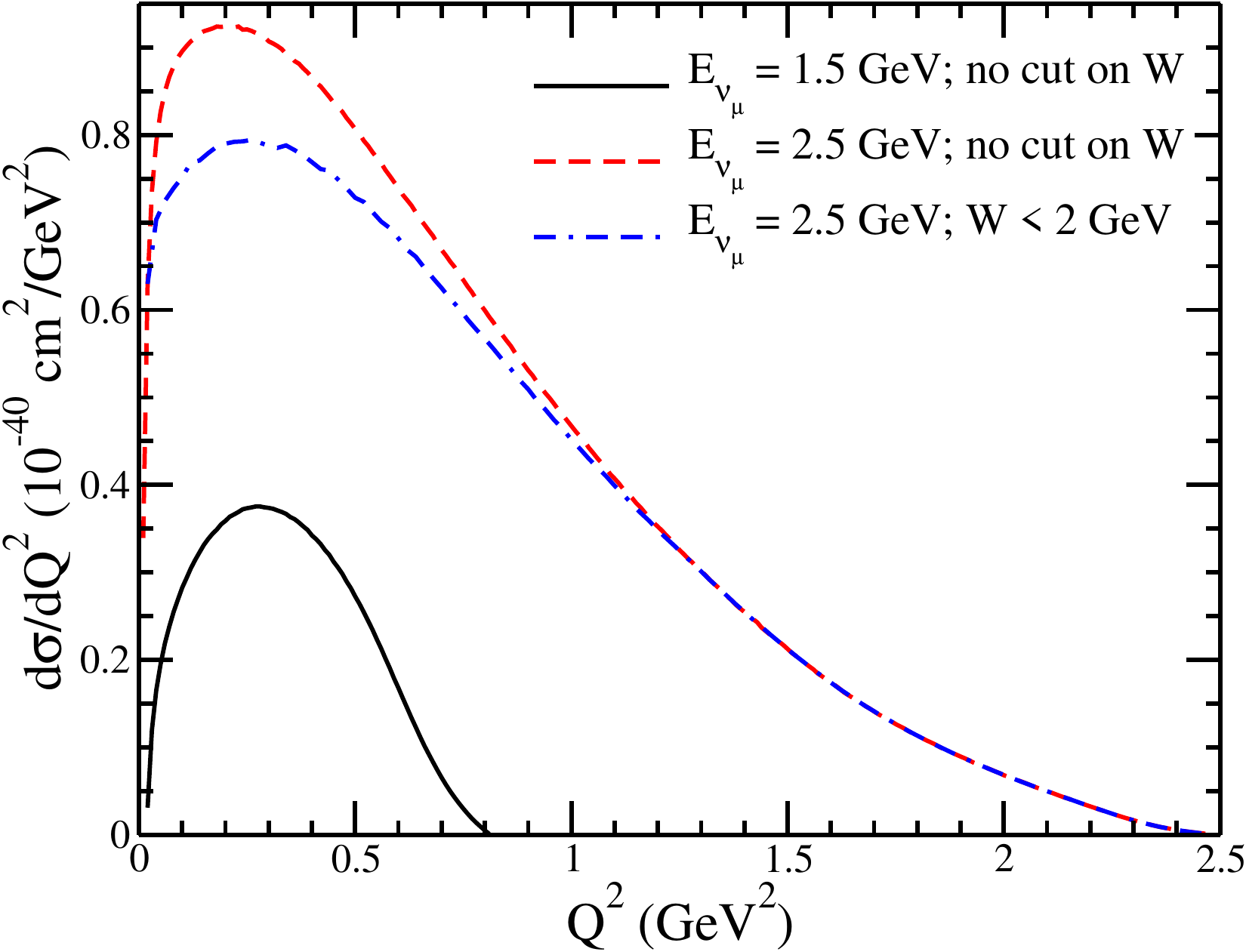}
\includegraphics[width=0.48\textwidth,height=8cm]{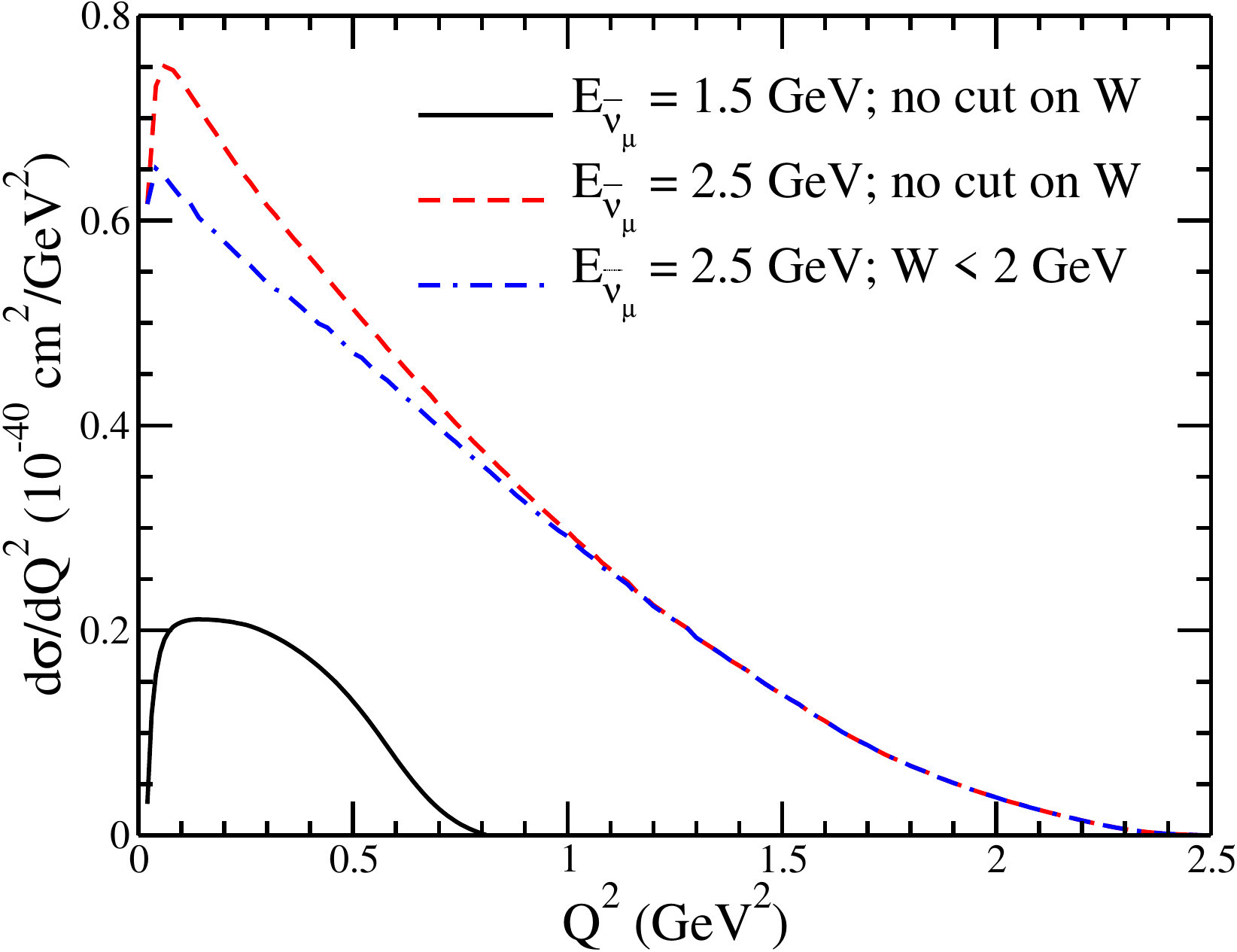}
\caption{$\frac{d\sigma}{dQ^2}$ vs. $Q^2$ for $\nu_\mu + n \rightarrow \mu^- + \Lambda + K^+$~(left panel) and $\bar\nu_\mu + p \rightarrow \mu^+ + \Lambda + K^0$~(right panel) processes at two different (anti)neutrino energies viz. 1.5~GeV~(lower curves) and 2.5~GeV~(upper curves). The solid line is the result when no cut on $W$ is applied for (anti)neutrino energy of 1.5~GeV, and the dashed line~(double dash-dotted line) is the result when no cut on $W$~(a cut of $W < 2$~GeV) is applied for (anti)neutrino energy of 2.5~GeV. }
\label{assoc-neut-q2}
\end{center}
\end{figure}

 \begin{figure}  
\begin{center}
\includegraphics[width=0.48\textwidth,height=8cm]{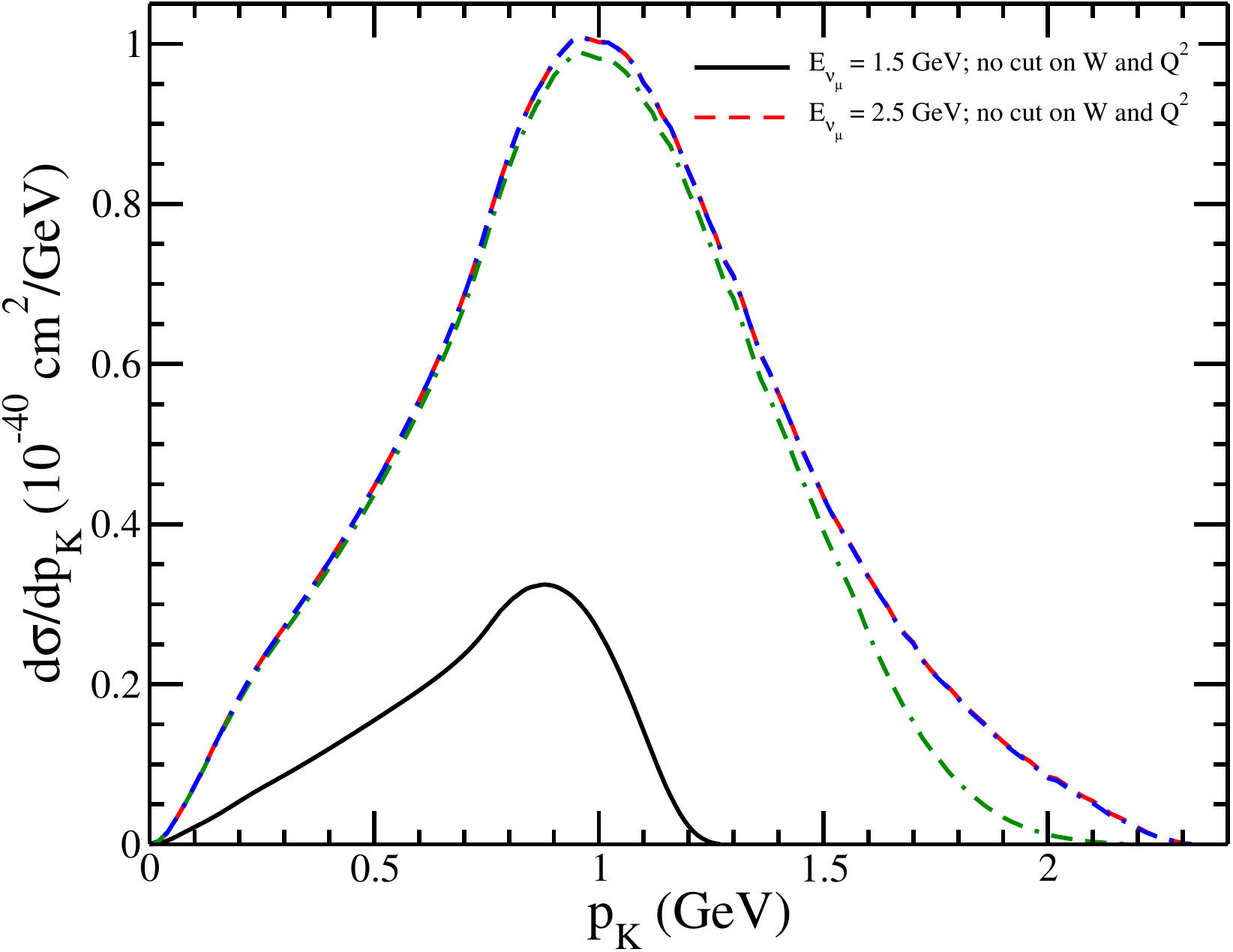}
\includegraphics[width=0.48\textwidth,height=8cm]{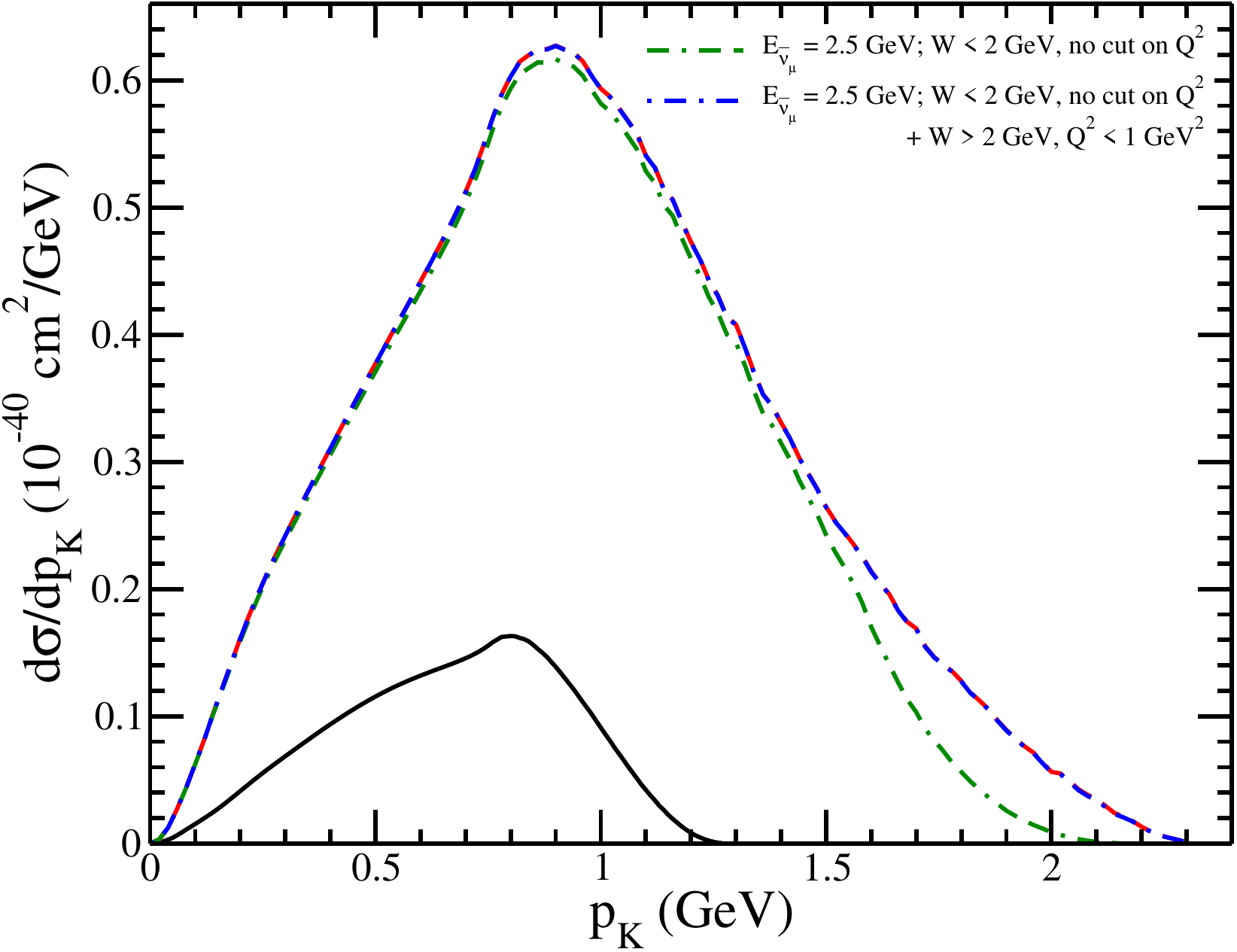}
\caption{$\frac{d\sigma}{dp_K}$ vs. $p_K$ for $\nu_\mu + n \rightarrow \mu^- + \Lambda + K^+$~(left panel) and $\bar\nu_\mu + p \rightarrow \mu^+ + \Lambda + K^0$~(right panel) processes. The solid line is the result when no cut on $W$ is applied for (anti)neutrino energy of 1.5~GeV, and the dashed line~(double dash-dotted line) is the result when no cut on $W$~(a cut of $W < 2$~GeV) is applied for (anti)neutrino energy of 2.5~GeV. The dash-dotted line represents the results at an (anti)neutrino energy of 2.5~GeV, when the cuts of $W<2$~GeV and $W\ge 2$~GeV with $Q^2<1$~GeV$^{2}$ are applied while evaluating the numerical calculations.}
\label{assoc-neut-pk}
\end{center}
\end{figure}

In Fig.~\ref{assoc-neut}, we have presented the results for the total cross section $\sigma$ for $\nu_\mu n \rightarrow \mu^- \Lambda K^+$~(left panel)  and $\bar\nu_\mu n \rightarrow \mu^+ \Lambda K^0$~(right panel)  processes as a function of incoming (anti)neutrino energy. These results have been presented when no cut on $Q^2$ and $W$ is applied, for the full model~(solid line) as well as for the individual contributions from several spin $\frac{1}{2}$ and $\frac{3}{2}$ resonances viz. $S_{11}(1650)$~(solid line with solid circles), $P_{11}(1710)$~(solid line with solid squares), $P_{13}(1720)$~(solid line with solid triangles), $P_{11}(1880)$~(dashed line), $S_{11}(1895)$~(double dash-dotted line), and $P_{13}(1900)$~(dash-double dotted line) resonances, considered in this work. For comparison, the results for the background contribution are also presented and depicted by the dotted line.  We find the contributions of $P_{11}(1710)$ and $P_{13}(1720)$ resonances as well as the background terms to be quite significant for both the neutrino as well as the antineutrino induced processes.
Quantitatively, in the case of neutrino induced $K\Lambda$ production channel, the $P_{11}(1710)$ resonance has a contribution of about 55\% at $E_{\nu_{\mu}}=1.5$~GeV, which decreases with the increase in neutrino energy and becomes 38\% at $E_{\nu_{\mu}}=2.5$~GeV. The $P_{13}(1720)$ resonance has a contribution of about 30\% in the energy range of $E_{\nu_{\mu}}=1.5-2.5$~GeV. While the nonresonant background terms have a contribution of about 17\% at $E_{\nu_{\mu}}=1.5$~GeV, which increases with the increase in neutrino energy and becomes 28\% at $E_{\nu_{\mu}}=2.5$~GeV. 
On the other hand, in the case of antineutrino induced reaction, the $P_{11}(1710)$ resonance has a contribution of about 40\% at $E_{\bar{\nu}_{\mu}}=1.5$~GeV, which decreases with the increase in antineutrino energy and becomes 34\% at $E_{\bar{\nu}_{\mu}}=2.5$~GeV. The $P_{13}(1720)$ resonance has a contribution of about 22\% at $E_{\bar{\nu}_{\mu}}=1.5$~GeV, which increases with the increase in antineutrino energy and becomes 27\% at $E_{\bar{\nu}_{\mu}}=2.5$~GeV. Whereas the nonresonant background terms have a contribution of about 20\% at $E_{\bar{\nu}_{\mu}}=1.5$~GeV, which increases with the increase in antineutrino energy and becomes 34\% at $E_{\bar{\nu}_{\mu}}=2.5$~GeV. 

In the few GeV energy region, the understanding of the transition from the hadronic regime, where hadrons are the effective degrees of freedom, to the deep inelastic regime, governed by the perturbative QCD dynamics of quarks and gluons, is of significant importance.
The intermediate region is commonly referred to as the shallow inelastic scattering~(SIS) region. 
However, defining the exact kinematic boundaries between SIS and DIS remains a challenge due to the limited theoretical understanding and a lack of high-statistics experimental data. 
Because of this scarcity of experimental data in the SIS region, studies of neutrino-nucleon and neutrino-nucleus scattering are primarily guided by the theoretical models. 
Athar and Morfin~\cite{SajjadAthar:2020nvy} have recently highlighted this issue, proposing the kinematic constraints of $W \ge 2$~GeV and $Q^2 \ge 1$~GeV$^{2}$ as cuts to distinguish contributions from the resonance and DIS regions. 
Nevertheless, defining a boundary between SIS and DIS is not straightforward as different experiments apply varying cuts on $W$, making comparisons across the datasets difficult. Following Ref.~\cite{SajjadAthar:2020nvy}, we have considered the region of $W<2$~GeV and all values of $Q^2$, as well as the region of $W>2$~GeV and $Q^2 < 1$~GeV$^{2}$ as the inelastic region, and obtained the results for the $K\Lambda$ production, while the region of $W \ge 2$~GeV and $Q^2 \ge 1$~GeV$^{2}$ has been excluded considering it to be purely the DIS regime.
{ However, the present choice on the numerical values of cuts on $W$ and $Q^2$ is motivated by the values used in the literature~\cite{SajjadAthar:2020nvy, Jeong:2023hwe} and more study is required to understand the transition region from resonance to DIS.}
In view of this, we have presented the results for the total cross section $\sigma$ for $\nu_\mu n \rightarrow \mu^- \Lambda K^+$~(left panel)  and $\bar\nu_\mu n \rightarrow \mu^+ \Lambda K^0$~(right panel)  processes as a function of incoming (anti)neutrino energy in Fig.~\ref{assoc-neut-cut}, by considering four different cases: 
\begin{itemize}
 \item [(i)] without applying any cuts on $W$ and $Q^2$~(shown by the solid line),
 \item [(ii)] when an upper cut of 2~GeV on $W$ is applied i.e. $W < 2$~GeV but without any cut on $Q^2$~(shown by the dashed line),
  \item [(iii)] when the cuts on $W$ and $Q^2$ are applied i.e. $W > 2$~GeV and $Q^2<1$~GeV$^{2}$~(shown by the dash-dotted line), and
 \item [(iv)] when a cut on $W <2$~GeV is applied without any cut on $Q^2$, and also included in this result are the contributions from  $W >2$~GeV having $Q^2 < 1$~GeV$^2$~(shown by the double-dash-dotted line).
\end{itemize}
   It may be observed from Fig.~\ref{assoc-neut-cut} that excluding the region of $W > 2$~GeV reduces the cross section for  $E_\nu > 2$~GeV. Whereas,  when the cross sections are evaluated by considering the contributions from $W < 2$~GeV with no cut on $Q^2$ and $W >2 $~GeV and $Q^2 < 1$~GeV$^2$, we find the results to be very close to the ones obtained by applying no constraints on $W$ and $Q^2$ i.e. the results shown by double dash-dotted line vs solid line are closed by  for the (anti)neutrino energies considered in this work.

In  Fig.~\ref{assoc-neut-compare}, we compare our results of the total scattering cross section with the old theoretical calculations and experimental results available in the energy region of a few GeV. The experimental data is from the BNL collaboration~\cite{Baker:1981tx} and the theoretical results are from  Shrock~\cite{Shrock:1975an}, obtained using the Born approximation and  Nakamura et al.~\cite{Nakamura:2015rta} obtained in the DCC model. 
Moreover, Shrock~\cite{Shrock:1975an} and Mecklenburg~\cite{Mecklenburg:1976pk} have also reported the results for the Argonne neutrino flux-averaged cross sections~($<\sigma>$) as 
$3.4 \times 10^{-42}$~cm$^{2}$ and $39.1 \times 10^{-42}$~cm$^{2}$, respectively, to be compared  with the experimentally derived data of the ANL experiment~\cite{Barish:1974ye}, i.e., $<\sigma> = 20 \times 10^{-42}$~cm$^{2}$.

In Fig.~\ref{assoc-neut-q2}, we present the results for the $Q^2$ distribution i.e. $\frac{d\sigma}{dQ^2}$ vs $Q^2$ at two representative (anti)neutrino energies viz. $E_\nu=1.5$~GeV~(lower curves) and 2.5~GeV~(upper curves) for $\nu_\mu n \rightarrow \mu^- \Lambda K^+$~(left panel)  and $\bar\nu_\mu n \rightarrow \mu^+ \Lambda K^0$~(right panel)  processes. These results are presented using the full contribution to the scattering amplitude i.e. resonant as well as nonresonant terms. These results are presented for the two different cases: (i) when no cut on $W$ is applied, and (ii) when a cut of 2~GeV on $W$ is applied i.e. $W < 2$~GeV. It may be observed from the figure that the effect of the $W$ cut is quite important in the peak region of $Q^2$, which decreases with the increase in $Q^2$ and becomes almost negligible at $Q^2 \ge 1$~GeV$^{2}$. 
{ The $Q^2$ distribution is more sharply peaked for the $K\Lambda$ production induced by the antineutrinos than for the neutrinos because of the different sign of the interference terms of the vector and axial vector form factors. 
The neutrino induced reactions occur with a constructive interference term which makes the cross section more pronounced in the low $Q^2$ region, while the antineutrino induced reactions occur with a destructive interference making the cross section more sharply peaked in the low $Q^2$ region.}

In Fig.~\ref{assoc-neut-pk}, we present the results for the kaon momentum distribution i.e. $\frac{d\sigma}{dp_K}$ vs $p_K$ at two representative (anti)neutrino energies viz. $E_\nu=1.5$~GeV~(lower curves) and 2.5~GeV~(upper curves) for $\nu_\mu n \rightarrow \mu^- \Lambda K^+$~(left panel)  and $\bar\nu_\mu n \rightarrow \mu^+ \Lambda K^0$~(right panel)  processes. These results are presented using the full contribution to the scattering amplitude i.e. resonant as well as nonresonant terms. These results are presented for the three different cases, one when no cut on $W$ and $Q^2$ is applied, second when a cut of 2~GeV on W is applied i.e. $W < 2$~GeV without applying any cut on $Q^2$, and the third one is by considering the case of $W < 2$~GeV and no cut on $Q^2$ along with the contribution when $W > 2$~GeV but $Q^2 < 1$~GeV$^2$. 
It may be observed that the results obtained by considering the combined effect of $W < 2$~GeV with no cut on $Q^2$ and $W>2$~GeV with $Q^2 < 1$~GeV$^2$ are almost comparable to the results obtained using no cut on $W$ and $Q^2$.

\section{Summary and conclusion}\label{summary}
We have studied the charged current $\nu_{\mu}(\bar{\nu}_{\mu})$ induced $K\Lambda$ production off the nucleon target in the few GeV energy region, which is relevant for the ongoing and proposed accelerator and atmospheric based neutrino experiments. 
The results are presented for the total scattering cross section and compared with the available theoretical and experimental works in the literature, as well as for the $Q^2$ distribution and the kaon momentum distribution.

The calculations have been performed using a model based on the effective Lagrangians to evaluate the contribution from the nonresonant and resonant diagrams. The contribution from the nonresonant background terms has been obtained 
using a microscopic model based on the SU(3) chiral Lagrangians. 
The contribution from the resonant diagrams due to the low lying spin $\frac{1}{2}$ resonances like $S_{11} (1650)$, $P_{11} (1710)$, $P_{11}(1880)$, $S_{11}(1895)$, and spin $\frac{3}{2}$ resonances like 
$P_{13} (1720)$ and $P_{13} (1900)$, for which there is finite branching ratios to the  $K\Lambda$ channel, has been considered using an effective phenomenological Lagrangian with its parameters determined from the experimental values of their branching ratios and decay widths to the $K\Lambda$ channel.
The vector form factors in the nucleon-resonance transition are expressed in terms of the electromagnetic $N-R$ form factors using the isospin symmetry, which in turn are determined in terms of the helicity amplitudes of the resonances.
The PCAC
hypothesis and generalized Goldberger-Treiman relation are used to fix the parameters of the axial-vector
current interaction. First the model has been used to study the photo- and electro- production of $K\Lambda$ and $\eta$ meson, to fix the vector part of the weak hadronic current.

We find that:
\begin{itemize}
 \item the present model explains the electromagnetic induced $\Lambda K$ and $\eta$ production data quite well. This is used to fix the various parameters of the vector current interaction in the (anti)neutrino induced associated particle production.
 
 \item the results of the total scattering cross section for (anti)neutrino induced $\Lambda K$ production off the nucleon target has contributions from the resonant and nonresonant background terms, which are both important. The resonance contribution is dominated by the $P_{11} (1710)$ and $P_{13} (1720)$ resonances.
 
 \item the effect of CM energy cut of $W < 2$~GeV reduces the cross section for $E_{\nu} > 2$~GeV, which is about 8\% at $E_{\nu}=2.5$~GeV and becomes 15\% at $E_{\nu}=3.5$~GeV, for both the neutrino and antineutrino induced $K\Lambda$ production. However, when the contribution of $W > 2$~GeV with $Q^2 < 1$~GeV$^2$ is taken into account, the numerical results are very close to the results obtained without any cut on $W$ and $Q^2$, for (anti)neutrino energies up to 3~GeV.
 
 \item $\frac{d\sigma}{dQ^2}$ peaks around $Q^2=0.2$~GeV$^2$ both for (anti)neutrino energies of 1.5~GeV and 2.5~GeV. The $Q^2$ distribution for the antineutrino induced $K\Lambda$ production is more sharply peaked as compared to the neutrino induced reaction.
 
 \item  the kaon momentum distribution $\frac{d\sigma}{dp_K}$ peaks around $p_{K}=$~(0.88)~1~GeV at $E_{\nu}=(1.5)~2.5$~GeV for the neutrino induced $K\Lambda$ production. The nature of kaon momentum distribution in the antineutrino induced reaction is almost similar to that observed in the neutrino case, except the fact that the peak shifts slightly to lower values of momentum in the case of antineutrino induced reaction.
\end{itemize}

To conclude, this is the first work which discusses $Q^2$ distribution and the kaon momentum distribution, and presents an updated calculation of the total scattering cross section for $\nu_\mu$ and $\bar\nu_\mu$ induced associated particle production of K$\Lambda$ from the free nucleon target. The energy region considered in this work may be quite useful in the experimental analysis of atmospheric and accelerator neutrino experiments like T2K, MicroBooNE, ICARUS, NOvA, MINERvA, SBND, and DUNE, where the average energy is expected to be around 2.5~GeV.

 \section*{Acknowledgments}
AF and MSA are thankful to the
Department of Science and Technology (DST), Government of India for providing financial assistance under Grant No.
SR/MF/PS-01/2016-AMU.

\end{document}